\documentclass{aastex63}
\usepackage{empheq}

\providecommand{\e}[1]{\ensuremath{\times 10^{#1}}}	
\providecommand{\R}{$R_{23}$}					
\providecommand{\etal}{{\it et al.} }			

\shorttitle{SFACT I}
\shortauthors{Salzer et al.}

\begin{document}


\title{The Star Formation Across Cosmic Time (SFACT) Survey.  I. Survey Description and Early Results from a New Narrow-Band Emission-Line Galaxy Survey}


\correspondingauthor{John J. Salzer}
\email{josalzer@indiana.edu}

\author[0000-0001-8483-603X]{John J. Salzer}
\affiliation{Department of Astronomy, Indiana University, 727 East Third Street, Bloomington, IN 47405, USA}

\author[0000-0002-4876-5382]{David J. Carr}
\affiliation{Department of Astronomy, Indiana University, 727 East Third Street, Bloomington, IN 47405, USA}

\author[0000-0002-5513-4773]{Jennifer Sieben}
\affiliation{Department of Astronomy, Indiana University, 727 East Third Street, Bloomington, IN 47405, USA}

\author[0000-0001-6776-2550]{Samantha W. Brunker}
\affiliation{Department of Astronomy, Indiana University, 727 East Third Street, Bloomington, IN 47405, USA}

\author[0000-0002-2954-8622]{Alec S. Hirschauer}
\affiliation{Department of Astronomy, Indiana University, 727 East Third Street, Bloomington, IN 47405, USA}
\affiliation{Space Telescope Science Institute, 3700 San Martin Dr., Baltimore, MD 21218, USA}

\begin{abstract} 
We introduce the {\it Star Formation Across Cosmic Time} (SFACT) survey.  SFACT is a new narrow-band survey for emission-line galaxies (ELGs) and QSOs being carried out using the wide-field imager on the WIYN 3.5 m telescope.  Because of the superior depth and excellent image quality afforded by WIYN, we routinely detect ELGs to r = 25.0.  Our survey observations are made using three custom narrow-band filters centered on 6590 \AA, 6950 \AA, and 7460 \AA.  Due to the sensitivity of the survey,  we are able to simultaneously detect sources via a number of different emission lines over a wide range of redshifts.  The principal lines detected in SFACT are H$\alpha$ (redshifts up to 0.144), [\ion{O}{3}]$\lambda$5007 (redshifts up to 0.500) and [\ion{O}{2}]$\lambda$3727 (redshifts up to 1.015).   In this paper we detail the properties of the survey as well as present initial results obtained by analyzing our three pilot-study fields.   These fields have yielded a total of 533 ELG candidates in an area of 1.50 deg$^2$ (surface density of 355 ELGs deg$^{-2}$).   Follow-up spectra for a subset of the ELG candidates are also presented.  One of the key attributes of the SFACT survey is that the ELGs are detected in discrete redshift windows that will allow us to robustly quantify the properties of the star-forming and AGN populations as a function of redshift to z = 1 and beyond.   
The planned acquisition of additional narrow-band filters will allow us to expand our survey to substantially higher redshifts.

\end{abstract}



\section{Introduction} 

Most of what is known about activity in galaxies has been learned by studying objects cataloged in dedicated surveys that probe for the telltale signs of that activity.  This is true regardless of whether that activity is due to above-average levels of star formation or is caused by accretion of matter onto a supermassive black hole.  These surveys have been carried out at wavelengths across the electromagnetic spectrum.  Early and extremely influential examples include the objective-prism survey for UV-excess galaxies carried out by Markarian and colleagues at the Byurakan Observatory \citep[e.g., ][]{mrk67, mrk81} and the 3C radio continuum survey carried out with the Cambridge radio interferometer \citep{edge1959, bennett1962, laing1983}.  

One of the most commonly adopted survey approaches for detecting activity has been to search for galaxies that exhibit strong optical or UV emission lines in their rest-frame spectra.  These emission-line galaxy (ELG) surveys have utilized a number of different selection techniques.   For example, several early ELG surveys utilized objective-prism spectroscopy to select their candidates  \citep[e.g.,][]{smith1975, UM1977, UM1981, Case82, Case83, Was83, SBS1, UCM94, UCM96, HS1999, HS2000, salzer2000, salzer2001, salzer2002}.  Alternatively, several surveys have been carried out using narrow-band (NB) imaging data \citep[e.g.,][]{BST1993, rw2004, Kakazu2007, werk2010, ly2011, hadot1, HIZELS12, HIZELS13, Stroe2015, cook2019, hadot2, LAGER20, hadot4, miniJPAS}.  Very strong-lined ELGs can be selected using standard broad-band (BB) colors when the line equivalent widths are very high \citep[e.g.,][]{rosenwasser2022}.  This latter approach has been successful in detecting extreme objects like the Green Peas \citep{gp} and Blueberries \citep{yang2017}.   Finally, the HETDEX survey \citep{hetdex2021} utilizes multiple integral field units to carry out a filled, non-targeted, wide-area ELG survey.

We have initiated a new NB survey for emission-line objects called SFACT: Star Formation Across Cosmic Time.  SFACT takes advantage of the instrumentation and excellent image quality of the WIYN 3.5 m telescope to deliver a rich and diverse catalog of star-forming galaxies from the local universe to z = 1, as well as AGN and QSOs to z $>$ 5. 

SFACT attempts to build upon the legacy of these previous ELG surveys.  It is being carried out using the NB imaging technique, and employs a unique set of custom filters that allows the survey to extend to high redshifts.  The high sensitivity of the telescope and camera combination allows us to be sensitive to multiple emission lines in each image, which provides a natural multiplexing component that makes the survey method more efficient.  

This paper presents a full description of the new NB ELG survey.  It also gives preliminary results from our pilot study,  including BB photometry, NB fluxes, and spectroscopic data for the sources detected in the first three survey fields.  In Section 2 we describe the motivations for carrying out the survey, while Section 3 lays out the overall survey design.  The results of our pilot study are given in \S 4, which includes example imaging and spectral data, summaries of the photometric properties of the SFACT candidates, and the presentation of their redshift, luminosity and star-formation rate (SFR) distributions.  We provide a comparison between SFACT and a number of recent NB surveys in \S 5 in order to highlight the similarities and differences of our new sample of ELGs relative to existing surveys.  Section 6 describes a number of planned applications for the survey, and includes additional example spectra to illustrate the utility of the survey for addressing a number of science questions.  The current status of the survey and future plans are presented in Section 7, and we summarize our study in \S 8.  

Two companion data papers present the survey results for our initial pilot study.   The first \citep[][hereafter SFACT2]{sfact2} presents the complete list of newly discovered SFACT candidates from our three pilot-study fields.  SFACT2 also includes more complete details of our observational methodology and data processing, as well as a full description of how our BB photometry and NB line-flux measurements are carried out.   The second \citep[][hereafter SFACT3]{sfact3} tabulates results from our initial follow-up spectroscopy of the pilot-study ELG candidates.   This paper describes in detail how our spectra are obtained and processed, and presents redshifts, line fluxes,  and key emission-line ratios.   Taken together, the three initial SFACT papers will serve to provide a complete description of our survey design, motivation, and methodology.   Subsequent papers in this series will present additional data sets, as well as results from science applications that are described below.

A standard $\Lambda$CDM cosmology with $\Omega_m = 0.27$, $\Omega_\Lambda = 0.73$, and $H_0 = 70$~kms$^{-1}$ Mpc$^{-1}$  is assumed in this paper.

\section{Motivation for the Survey} 

Several factors have driven the  development of the SFACT survey, both scientific and practical.  In this section we briefly describe the factors that motivated us to carry out this project, since they serve to shape the survey design and methodology.

The combination of WIYN 3.5 m telescope\footnote{The WIYN Observatory is a joint partnership of the University of Wisconsin-Madison, Indiana University, Pennsylvania State University, Purdue University, University of California  - Irvine, and the NSF's NOIRLab.} and the wide field One Degree Imager (ODI)  camera offers an almost unique opportunity to carry out the SFACT survey.   As described in the next section, WIYN and its instrumentation suite allow us to execute the SFACT survey in a way that plays to both the strengths of the facility and to the strengths of the science goals of the project.   In particular, our ability to use the same telecsope with a multi-object spectroscopic instrument efficiently provides for essential follow-up spectroscopy of our candidate ELGs. 

Naturally, a key motivation for carrying out SFACT revolves around the science goals of the project.  The planned science applications are broad, covering a range of topics in extragalactic astronomy.   The namesake project of the survey is to carefully measure the star-formation rate density of the universe from z = 0 to z = 1 and beyond.   In addition, we also plan to evaluate the evolution of the metal abundances of galaxies over a similar redshift range.  We will use SFACT to detect and characterize the population of AGNs over the redshift range covered by the survey, and to probe the evolution of their volume densities and metallicities with lookback time.  We will explore the environments of all of our ELG populations, both star-forming and AGN, by utilizing deep redshift survey information carried out in each SFACT field.  We will discover numerous examples of extreme but rare objects at a wide range of redshifts: Green Peas \citep{gp, brunker2020}, Blueberries \citep{yang2017}, and extremely metal-poor dwarf galaxies \citep[e.g.,][]{hirschauer2016, mcquinn2020}.  Finally, the comprehensive nature of our survey will allow us to place these extreme types of galaxies into context with the overall population of star-forming galaxies at each redshift covered by the survey.  Section 6 gives a more comprehensive discussion of each of these science applications that are planned for SFACT.

The SFACT survey is, in many ways, an outgrowth of the H$\alpha$ Dot survey \citep{hadot1, hadot2, hadot4}.  As an additional motivation, we mention the desire to build upon (and improve upon) this  and other previous NB surveys.   The SFACT program adds a number of new aspects that makes it fairly unique.   Based on the observed depth of the H$\alpha$ Dot survey presented in \citep{hadot4}, we are able to predict the depth of the putative SFACT survey.   Accounting for the differences in the telescope apertures and filter widths, we arrive at an estimate for the new survey to reach a median NB emission-line flux of $\sim$2 $\times$ 10$^{-16}$ erg s$^{-1}$ cm$^{-2}$.   Furthermore, we predict a typical surface density of $\sim$80-120 emission-line objects per filter per deg$^{2}$.  By utilizing multiple NB filters on the same fields, one naturally achieves a multiplexing advantage.  With the current set of three NB filters, we expect to discover 240-360 ELGs deg$^{-2}$, or between 120 and 180 objects per field.   Since this large number of ELGs are detected in a volume limited by the NB filters, the projected volume densities of star-forming galaxies and AGN will be quite high.  The promise of achieving these large volume densities was a strong motivator for carrying out the SFACT survey.

\section{Design of the Survey} 

\subsection{Telescope and Instrumentation}

The concept for the SFACT survey was developed around the WIYN 3.5 m telescope and the ODI camera.   Several important factors drove the survey design.   

\noindent (i) The large aperture and superior image quality delivered by the WIYN telescope will allow the survey to reach an excellent depth.   Based on our previous experiences with NB imaging on smaller telescopes with the H$\alpha$ Dot survey (see \S 2), we projected that we would be able to achieve a median r-band magnitude of approximately 22.7 for the survey, and to detect faint emission-line objects to r $\sim$ 25.0.  

\noindent (ii) The f/6.3 beam of the Nasmyth focus on WIYN is slow enough to allow for use with NB filters.  Typically, wide-field imagers are fed by fast beams which, when used with NB filters, result in strong spatial variations of the effective transmitted bandpass on the detecter.  The slower convergence of the WIYN Nasmyth focus provides for a more uniform bandpass across the detector.  

\noindent (iii) The advent of the ODI camera.  The camera has an image scale of 0.11\arcsec\ pixel$^{-1}$, which allows it to take advantage of the excellent image quality delivered by the WIYN telescope.  ODI was originally designed to deliver a full 1$^\circ$ $\times$ 1$^\circ$ FOV,  
However, the final commissioned version of ODI has a FOV of 48$^\prime$ $\times$ 40$^\prime$, for a survey area of $\sim$0.53 sq. deg..  
The loss of nearly half of the expected detector area is unfortunate, as
it necessitated a longer time frame for carrying out the program, essentially doubling the project timescale.

\noindent (iv) The availability of the Hydra multi-fiber positioner on WIYN allows for a direct path for acquiring follow-up spectra of our candidates.  These spectra play a central role in the use of the SFACT ELG sample for carrying out the various science applications planned for the survey constituents (e.g., \S 6).  
The emission-line nature of the SFACT candidates translates into our ability to use WIYN for {\it both} the imaging selection and the spectroscopic confirmation.

\noindent (v) The recognition that, at the photometric depth achievable with WIYN, we would be sensitive to multiple emission lines from ELGs and QSOs present in each survey field.  That is, a NB filter at a fixed wavelength would simultaneously be sensitive to H$\alpha$ emission from low-redshift galaxies, to the [\ion{O}{3}]$\lambda$5007 line at intermediate redshifts, to [\ion{O}{2}]$\lambda$3727 at higher redshifts, and to the various UV lines from QSOs (e.g., \ion{Mg}{2} $\lambda$2798, \ion{C}{3}] $\lambda$1909, \ion{C}{4} $\lambda$1548, Ly$\alpha$ $\lambda$1215) at high redshifts.   The concept from the start was to utilize a limited number of NB filters and to think of the survey in terms of probing specific {\it redshift windows}, with different emission lines being detected in different windows.  The concept of redshift windows will be a central theme throughout the rest of this paper.

\subsection{Narrow-Band Filters}

The specific choice of the SFACT NB filters represents a combination of compromise and scientific opportunity.  Among the key drivers for the selection of the filters were their size and cost, plus the fact that ODI filter assembly is only capable of holding nine filters at a time.  Five of these filter slots are permanently designated to hold the standard {\it ugriz} BB filters.  This leaves only four slots available for any additional filters.  The full-field ODI filters are 42 $\times$ 42 cm, and cost in the neighborhood of \$70-\$80k each.   These two factors immediately suggest that creating a large set of NB filters with contiguous and overlapping wavelength coverage is not a realistic goal.    

Due to the high cost and uncertainties concerning the fabrication of such large NB filters, we opted to initially purchase a single NB filter.  
After much internal debate, we settled on a central wavelength of 6950 \AA.  The width of the filter was set at $\sim$90 \AA, driven largely by the concerns of the potential vendors over meeting design specifications if the bandpass were too narrow.  For the purpose of our survey, this bandwidth represents a compromise.  In NB surveys such as ours, a smaller bandwidth will result in higher sensitivity for emission lines, since it reduces the diluting effect of the underlying continuum within the bandpass (i.e., results in a higher contrast for the line).  For this reason, most extragalactic H$\alpha$ filters sets are designed with bandwidths of 50 -- 60 \AA\ or smaller \citep[e.g.,][]{vansistine2016}.   For example, the filters used to identify planetary nebulae in nearby galaxies \citep[e.g.,][]{jacoby1990} had bandwidths of $\sim$30 \AA.  The use of $\sim$90 \AA\ wide filters will result is a somewhat lower sensitivity for SFACT.  On the other hand, the larger bandwidth of the SFACT filters results in a correspondingly larger redshift range over which ELGs can be detected.  This increases the survey volume and hence the number of objects that are detected.

\begin{deluxetable*}{ccccccc}
\tabletypesize{\footnotesize}
\tablewidth{0pt}
\tablecaption{Properties of the SFACT NB Filters}
\label{tab:filters}

\tablehead{
 \colhead{Filter} &  \colhead{$\lambda_{cent}$} &  \colhead{$\Delta\lambda$} &  \colhead{z range - H$\alpha$} &\colhead{z range - [\ion{O}{3}]} &  \colhead{z range - [\ion{O}{2}]} &  \colhead{z range - Ly$\alpha$}  \\
  &  \colhead{\AA} &  \colhead{\AA} &  &  &  &   \\
 (1)  & (2)  & (3)  & (4)  & (5)  & (6)  & (7) 
}
 
\startdata
 NB659 = NB2 & 6590 &  81.1 &  -0.002 -- 0.011 & 0.308 -- 0.325 & 0.757 -- 0.780 & 4.389 -- 4.458  \\
 NB695 = NB1 & 6950 &  91.0 & \ 0.052 -- 0.066 & 0.378 -- 0.397 & 0.852 -- 0.877 & 4.680 -- 4.759  \\
 NB746 = NB3 & 7460 &  96.7 & \ 0.129 -- 0.144 & 0.480 -- 0.500 & 0.988 -- 1.015 & 5.099 -- 5.181  \\
 \ NB812* & 8120 &  90 & \ 0.230 -- 0.244 & 0.613 -- 0.631 & 1.166 -- 1.191 & 5.646 -- 5.720  \\
 \ NB912* & 9120 &  90 & \ 0.383 -- 0.397 & 0.812 -- 0.830 & 1.435 -- 1.459 & 6.469 -- 6.543 
\enddata
\tablecomments{ *\ Proposed additional SFACT filters}
\end{deluxetable*}

After the delivery of the first survey filter in June 2016 (NB695, designated as NB1 within the SFACT survey), we carried out a series of test observations.  The results of these preliminary observations verified the validity of our survey method.  Once this was established, we proceeded to order two additional NB filters for ODI.  The first of these was selected to have a central wavelength of 6590 \AA\ (NB659, designated as NB2), and could serve as a zero redshift H$\alpha$ filter for general purpose use in addition to being useful for SFACT.  The second has a central wavelength of 7460 \AA\ (NB746, designated as NB3).  Table~\ref{tab:filters} lists the key features for all three ODI NB filters, including the central wavelength ($\lambda_{cent}$), the filter width ($\Delta\lambda$, defined as the full width at half the maximum transmission level), and the redshift ranges covered by each filter for several strong emission lines.  These new filters were delivered in August 2018 and were put into service immediately.   For all observing runs from Fall 2018 onward, the full suite of three NB filters have been used to observe all survey fields.

\begin{figure}
\centering
\includegraphics[width=3.5in]{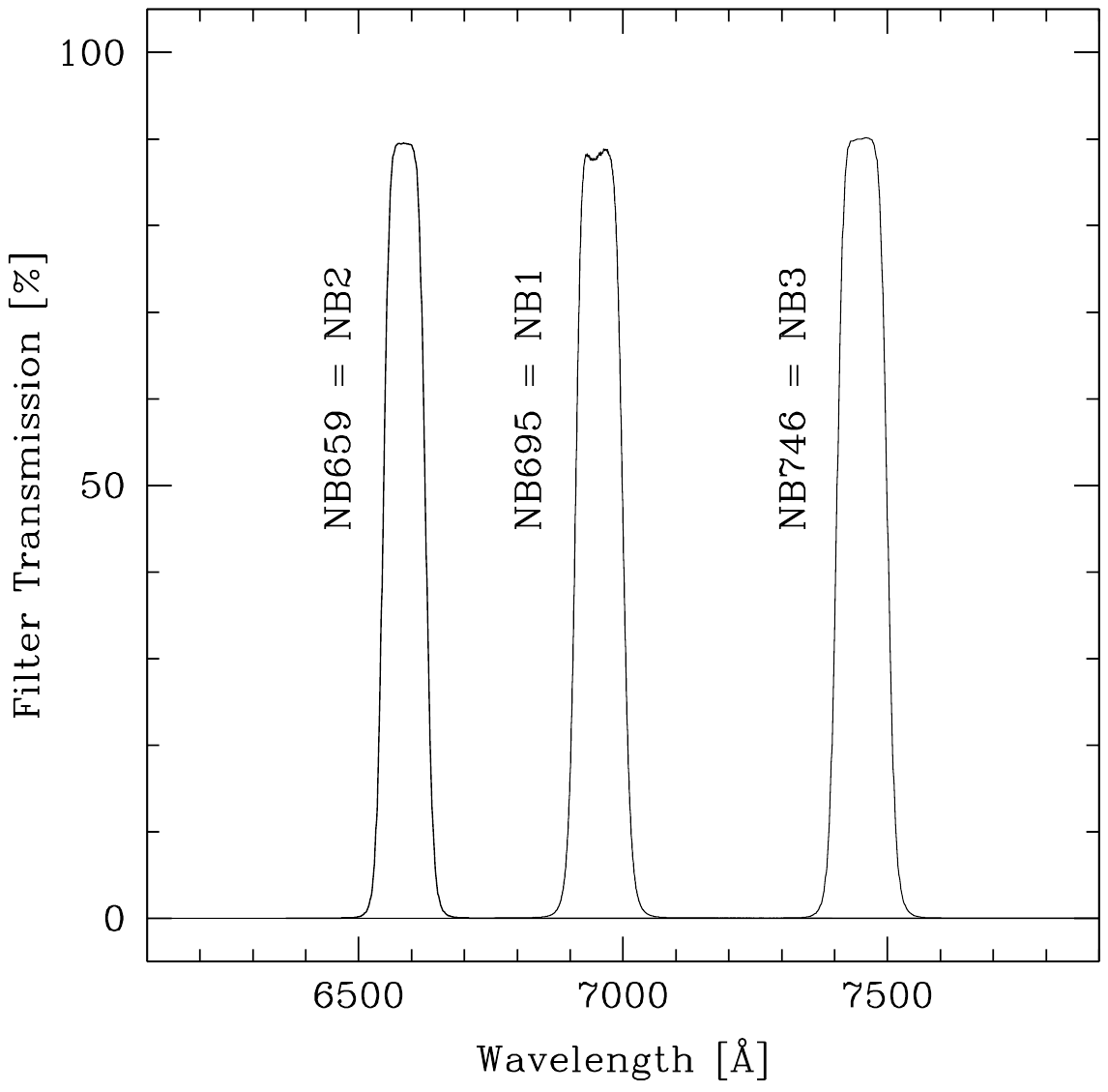} \includegraphics[width=3.45in]{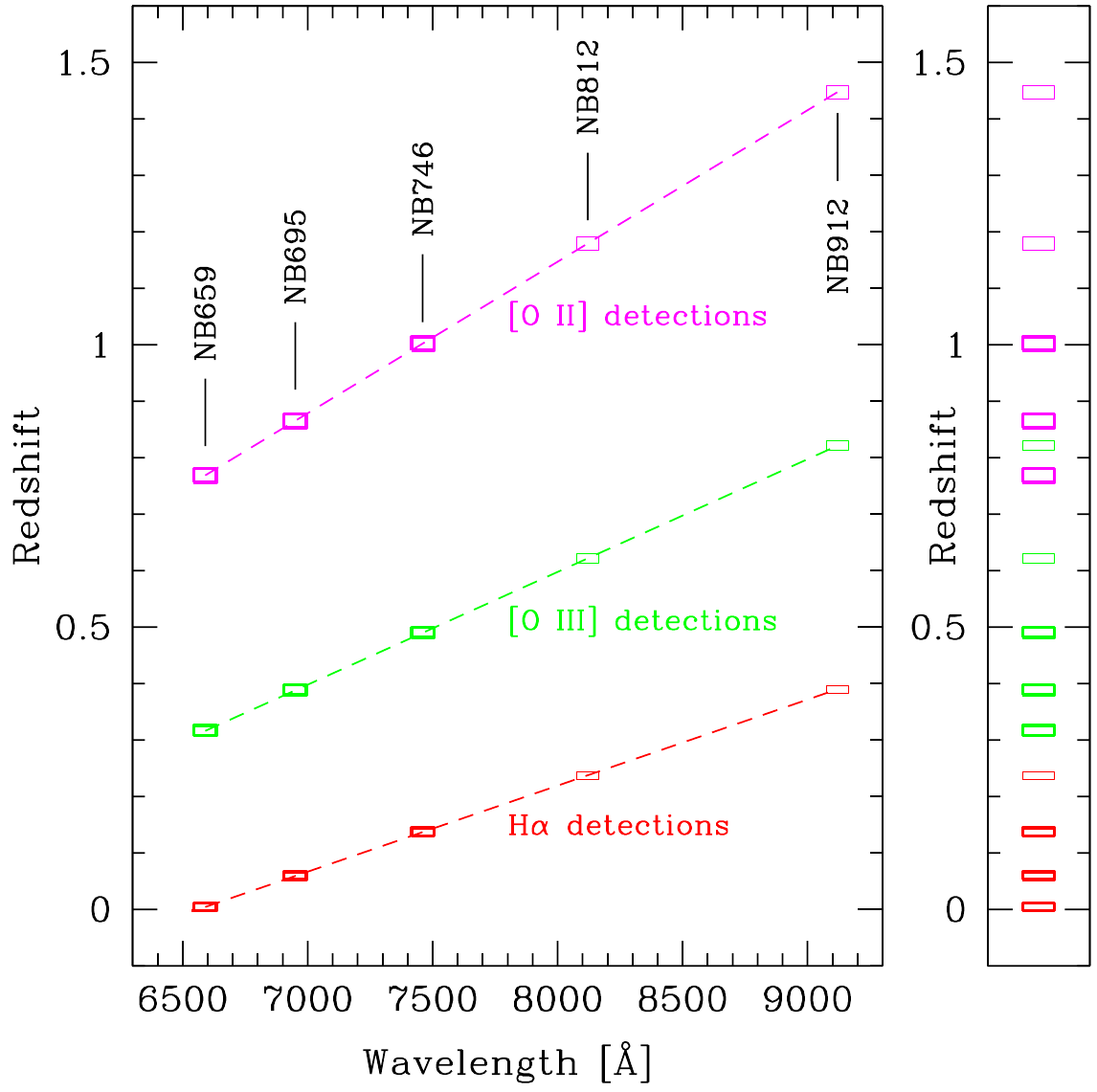}

\caption{{\it Left}: Filter tracings of the three NB filters currently in use with SFACT.  {\it Right}: Plot showing the wavelength coverage of the SFACT NB filters as well as the redshift ranges associated with the detection of strong optical nebular emission lines.  Each vertical column represents a specific NB filter, while the vertical location of each rectangular box represents the redshift range for the emission line indicated.  We show the redshift windows only for H$\alpha$, [\ion{O}{3}]$\lambda$5007, and [\ion{O}{2}]$\lambda$3727; the majority of the SFACT candidates are detected via one of these three lines.  Also shown are the two additional filters that we plan to add to the survey.  The plot on the right-hand side compresses the wavelength scale in order to better illustrate the distribution of the redshift windows covered by the survey.}
\label{fig:filters}
\end{figure}

Figure~\ref{fig:filters} shows the transmission profiles of the three SFACT NB filters, as well as the redshift coverage for the key optical lines of H$\alpha$, [\ion{O}{3}]$\lambda$5007, and [\ion{O}{2}]$\lambda$3727.  In particular, the righthand portion of the figure illustrates the redshift windows surveyed by our method.  The subplot on the far right suppresses the wavelength range and shows the spacing of the redshift windows.  As can be seen, the wavelengths of the filters were chosen to yield a fairly uniform redshift coverage for these windows.   The figure includes two additional filters that are planned for as part of a future expansion for the survey.  These latter two filters are located in the well known gaps of the telluric OH spectral lines at $\sim$8120 \AA\ and $\sim$9120 \AA.  These two new filters will help to fill in the gaps in the current distribution of redshift windows, as well as extend the overall redshift coverage to redshifts approaching 1.5 for the strong optical nebular lines.   Their proposed properties are also listed in Table~\ref{tab:filters}.

We point out two additional features regarding the SFACT NB filters.  First, the location of the NB695 filter was 
selected such that galaxies detected with it via their [\ion{O}{3}]$\lambda$5007 emission would have their H$\alpha$ + [\ion{N}{2}] lines redshifted into the night sky OH gap at $\sim$9120 \AA.   Hence objects detected via [\ion{O}{3}] emission in NB695 will have H$\alpha$ emission falling in NB912 (see Table~\ref{tab:filters}).   Second, the wavelength range of NB659 is such that only extremely low-redshift systems will be detected via their H$\alpha$ line (z $<$ 0.010).   This in turns means that the volume surveyed with this filter for H$\alpha$ emitters will be very small.  Hence, very few H$\alpha$-detected galaxies are expected to be found with this filter.  There is only one such object in the three pilot study fields presented in the current paper, 
a very low-luminosity dwarf star-forming galaxy.

\vfill\eject
\subsection{Planned Survey Size and Location of Survey Fields}

One of the goals of the SFACT survey is to deliver large enough samples of ELGs to be able to carry out robust derivations of astrophysical interest within {\it each redshift window}.  Examples of these derivable products include emission-line selected luminosity functions, star-formation rate densities, and mass-metallicity/luminosity-metallicity relations.  Hence, we established a goal of detecting 800-1200 ELGs within each redshift window when summed over all survey fields.  

As indicated in \S 2, the expectation for the typical number of ELGs detected per filter per field is 40-60 objects.  This will of course vary substantially from field-to-field due to cosmic variance.  Using this number as an average per field, and assuming that the three survey filters will yield similar numbers of detections when averaged over all of the survey fields, one arrives at an estimate of the number of fields that would be required to complete the survey: 50 to 60 fields.  This represents 25 - 30 deg$^2$ of sky coverage.  When completed, the SFACT survey will catalog between 6000 and 11,000 ELGs using the current three survey filters, and proportionally more when the two additional filters (NB812 and NB912) are added.

The selection of the SFACT survey field locations follows a number of criteria.   Fields need to be located within the footprint of the Sloan Digital Sky Survey \citep[SDSS; ][]{sdss} to facilitate the photometric calibration of our images (see SFACT2).  Most fields are located at high Galactic latitude to minimize foreground extinction, and have declinations between roughly 10$^\circ$ and 50$^\circ$ so that they transit within 20$^\circ$ of the zenith from Kitt Peak.  Since the total survey area is fairly small, we opted to observe a series of widely-scattered fields in both the Spring and Fall observing seasons to help ensure that the overall survey is not subject to cosmic variance (i.e., highs and lows in the galaxy density).  This latter criterion is especially important for SFACT, since the NB filters only cover limited swaths of redshift space.  It is easy to imagine that in some fields one or more of the filters may be surveying a low density void at the redshift of one of the key emission lines and yield very few detections.  Conversely, in some fields the same filters may hit a rich galaxy filament or cluster environment, resulting is an excess of detections.  By averaging over several dozen widely-spaced fields, we expect to completely wash out the effects of cosmic variance.

Many of the SFACT field locations observed during the early stages of the project are centered on known Green Pea galaxies \citep[e.g.,][]{gp, brunker2020}.  These extreme star-forming galaxies were selected from [\ion{O}{3}]-detected ELGs in either the KISS \citep{salzer2000, salzer2001, gronwall2004, jangren2005} or H$\alpha$ Dot \citep{hadot1, hadot2} surveys.  The reasons for this are two-fold.  First, the locations of these Green Peas are distributed fairly randomly across  broad areas of both Galactic caps.  This satisfies one of the criteria specified above.  Second, these fields have been the subject of a focused redshift survey \citep[e.g.,][]{brunker2022}.
The data from this redshift survey will provide a deep comparison sample for future studies that look into the impact of the local environment on the properties of the SFACT ELGs.

Additional SFACT field locations have been selected based simply on the availability of observing time, the desire to observe fields that are widely spaced across the sky, and the need to observe within $\sim$3 hours of the meridian.  If no suitable Green Pea target is available within a range of right ascension that is up during a scheduled observing run, we select an appropriate field location that is devoid of bright stars that allowed us to fill our observing schedule.

\section{Preliminary Results from the Pilot Study} 

The first observing run during which the two newer NB filters (NB2 and NB3) were available took place in September 2018.  During this run we completed the observations for three fields with all six filters (gri broad-band plus NB1, NB2 and NB3 narrow-band; see SFACT2) for the first time.   These three fields were designated as the SFACT pilot-study fields.  The data obtained from our pilot-study observations were used to test our analysis methods and to fine-tune our selection software.  They are also the first fields for which substantial follow-up spectroscopy was obtained.   In this section we present initial results from our analysis of these three fields.

Basic information about the three pilot-study fields is given in Table~\ref{tab:pilot_study}.  This includes the field designation, where we adopt the nomenclature SFF\#\#.  Here SFF stands for SFACT Fall (for Fall fields); Spring fields are designated SFS\#\#.  The number is a running number that designates each field within a given season.   For the SFF fields the first fifteen field locations are listed in ascending RA order, while subsequent fields are numbered in the order in which the imaging observations are completed.  The remaining information presented in the table includes the name of the Green Pea galaxy that the field is centered on (if appropriate), the celestial coordinates of the field center, the number of ELG candidates detected in each of the three NB filters, the total number of SFACT candidates in each field, and the number for which follow-up spectroscopy currently exists.

\subsection{Results from the Imaging Survey}

In this section we illustrate the observationally derived properties of the SFACT ELG candidates utilizing the results from our pilot study.  The details of the observations, object selection, and photometric analysis are presented in the SFACT2 companion paper.  Here we give a brief description of how our ELG candidates are selected, to help give context to the subsequent presentation describing the properties of our survey constituents.

The procedure for selecting SFACT objects follows common practice \citep[e.g.,][]{hadot1, HIZELS12, HIZELS13, Stroe2015, cook2019, hadot2, LAGER20, hadot4, miniJPAS}.   In particular, we follow the methodology developed for the H$\alpha$ Dots survey \citep{hadot1, hadot2, hadot4}.  An automated object-finding routine catalogs every source located within each of our survey fields.  Next we measure the fluxes for every object in each of the three NB images as well as each of the corresponding ``continuum" images (constructed from the BB images).  We then identify the objects that possess a statistically significant excess of NB flux, and flag them as ELG candidates.   Specifically, we consider objects as ELG candidates if they display an excess of flux in the NB image that is 0.4 magnitudes brighter than the flux in the continuum image {\it and} if the detected excess is at or above the 5$\sigma$ level.   See SFACT2 for details.

\subsubsection{Number of Detected Emission-Line Objects}

The total number of emission-line detections for the pilot study fields is 533 (see Table~\ref{tab:pilot_study}).  Since the area covered by these three fields is 1.5025 deg$^2$, the surface density of objects detected is 355 ELGs deg$^{-2}$.  This number is consistent with the projected surface density estimate derived in \S 2.  As alluded to in \S 3, the number of ELGs detected in a given field through a given filter is highly variable, being dependent on the large-scale structure present in each field (i.e., cosmic variance).   For example, the number of ELGs detected in SFF10 varies by a factor of $\sim$5 (22 objects in NB2, and 110 in NB3).   The implication is that the NB2 filter is sampling mostly low-density regions in this direction, while both NB1 and NB3 are intersecting some high-density portions of the universe.  The other two fields show substantially lower variations between filters, but there is significant variation between the fields (e.g., 132 ELGs for SFF01, and 216 for SFF10).

\begin{deluxetable*}{ccccccccc}
\tabletypesize{\footnotesize}
\tablewidth{0pt}
\tablecaption{The SFACT Pilot-Study Fields}
\label{tab:pilot_study}

\tablehead{
 \colhead{ } & \colhead{Central} & & & \colhead{} &\colhead{\# Detections} & & \colhead{Total \# } & \colhead{\# with} \\
 \colhead{Field} & \colhead{Object} &  \colhead{$\alpha$(J2000)} &  \colhead{$\delta$(J2000)}   & NB1  & NB2 & NB3 & \colhead{Candidates } & \colhead{Spectra} \\
  (1)  & (2)  & (3)  & (4)  & (5)  & (6) & (7) & (8) & (9)
}
\startdata
SFF01 & --- & 21:43:00.0 &  20:00:00.0 & \ 41 & \ 47 & \ 44  & 132 & 110\\
SFF10 & H$\alpha$ Dot 55 & \ 1:44:40.3 &  27:54:35.5 & \ 84 & \ 22 & 110 & 216 & 201 \\
SFF15 & H$\alpha$ Dot 22 & \ 2:39:12.6 &  27:52:02.3 & \ 67 & \ 71 & \ 47 & 185 & 146
\enddata
\end{deluxetable*}

The average number of objects detected per filter per field are: NB1 = 64.0, NB2 = 46.7, NB3 = 67.0.  While these averages are still subject to significant uncertainty due to cosmic variance between the three survey fields, they reflect the expectation indicated in \S 3 that the number of objects detected in NB2 will be lower than the other two filters due to the fact that the survey volume for low-redshift objects (H$\alpha$ detections) will be extremely small.  A naive expectation would be that there will be essentially zero H$\alpha$ detections with NB2, leading to a typical detection rate with that filter that is $\sim$2/3rds the value for NB1 and NB3.  Our preliminary numbers are approximately consistent with that expectation.

As alluded to in SFACT2, our object selection for the pilot-study fields was meant to be inclusive, in the sense that we tended to include a number of more questionable sources in the final catalogs.  This is not to say that we relaxed the quantitative selection criteria for these fields.   Rather, we were more inclusive during the final visual inspection of the candidates where we identify and reject false detections such as image artifacts.  The motivation for doing this was to include a number of the more questionable objects in our spectroscopic follow-up lists, with the plan to let the spectra verify or reject these more dubious sources.  By using the follow-up spectra as a guide, we better trained ourselves to identify the false detections in the subsequent survey fields.  For this reason, the number of cataloged sources in these three fields is likely to be higher than we might typically expect in future survey lists, and the fraction of true emission-line objects is likely to be lower.

\subsubsection{Categories of Emission-Line Objects Cataloged}

The SFACT survey is designed to detect emission lines in a broad range of extragalactic sources.  The varied nature of the sources selected require us to differentiate between classes of objects, exclusively for the purpose of expediting the photometric measurements and the follow-up spectroscopy.  We currently recognize three categories of objects: (1) compact objects with a centralized emission region, (2) extended galaxies with one or more \ion{H}{2} regions, and (3) bright \ion{H}{2} regions.  We stress that these three classes are used only to differentiate how the objects are treated by the SFACT analysis and measurement software; they do not represent distinct types of ELGs (e.g., starburst nuclei, Seyferts, QSOs, etc.).  The three categories of SFACT objects are illustrated in Figure~\ref{fig:images1} below.

The vast majority of the SFACT emission-line objects fall into the first group of compact objects.   This includes essentially all of the objects that are detected via their [\ion{O}{3}] line (redshifts of 0.30 - 0.50), all of the [\ion{O}{2}] detections (redshifts 0.75 - 1.02), and QSOs.   It also includes the smaller, more compact H$\alpha$ detections.   Objects in this class have the bulk of their emission emanating from the central region of the galaxy.  This allows us to use the photometric center of the object as the location for measuring both the BB magnitudes as well as the emission-line flux.   The same location is used for follow-up spectroscopy.

The survey often detects individual \ion{H}{2} regions in nearby galaxies (H$\alpha$ detections, redshifts of 0.00 - 0.15).  In some cases, large, face-on spirals yield dozens of \ion{H}{2} regions.   While we retain the full set of individual \ion{H}{2} regions in our internal lists, for the purposes of the survey we do not include all of the \ion{H}{2} regions in our final catalog.  Rather, we record the location of the galaxy center, so that the BB magnitudes and {\it total} emission-line flux from the galaxy can be accurately measured.  These are the category 2 objects.  

In some cases, there is no detectable emission associated with the galaxy center of the category 2 objects.  For this reason, we typically also catalog the single brightest \ion{H}{2} region in the extended galaxies and target it for follow-up spectroscopy.  These are the category 3 objects.  This scheme ensures that the brightest emission region in each SFACT detection is observed spectroscopically.  Objects in this group are always a sub-unit of an object in category 2, although not all objects in category 2 have a corresponding \ion{H}{2} region in our final catalog.  The latter circumstance will occur when, for example, the center of the galaxy possesses a strong emission source.  

The number of detections listed in Table~\ref{tab:pilot_study} do not include all of the individual \ion{H}{2} regions detected by our software.  Rather, each detected {\it galaxy} appears only once, unless there is a bright \ion{H}{2} region (category 3) associated with it.   In our pilot-study catalogs, there are 40 extended galaxies (category 2) and 19 \ion{H}{2} regions contained in the three fields.

\subsubsection{Broad-band Photometry of the Candidates}

\begin{figure}
\centering
\includegraphics[width=3.35in]{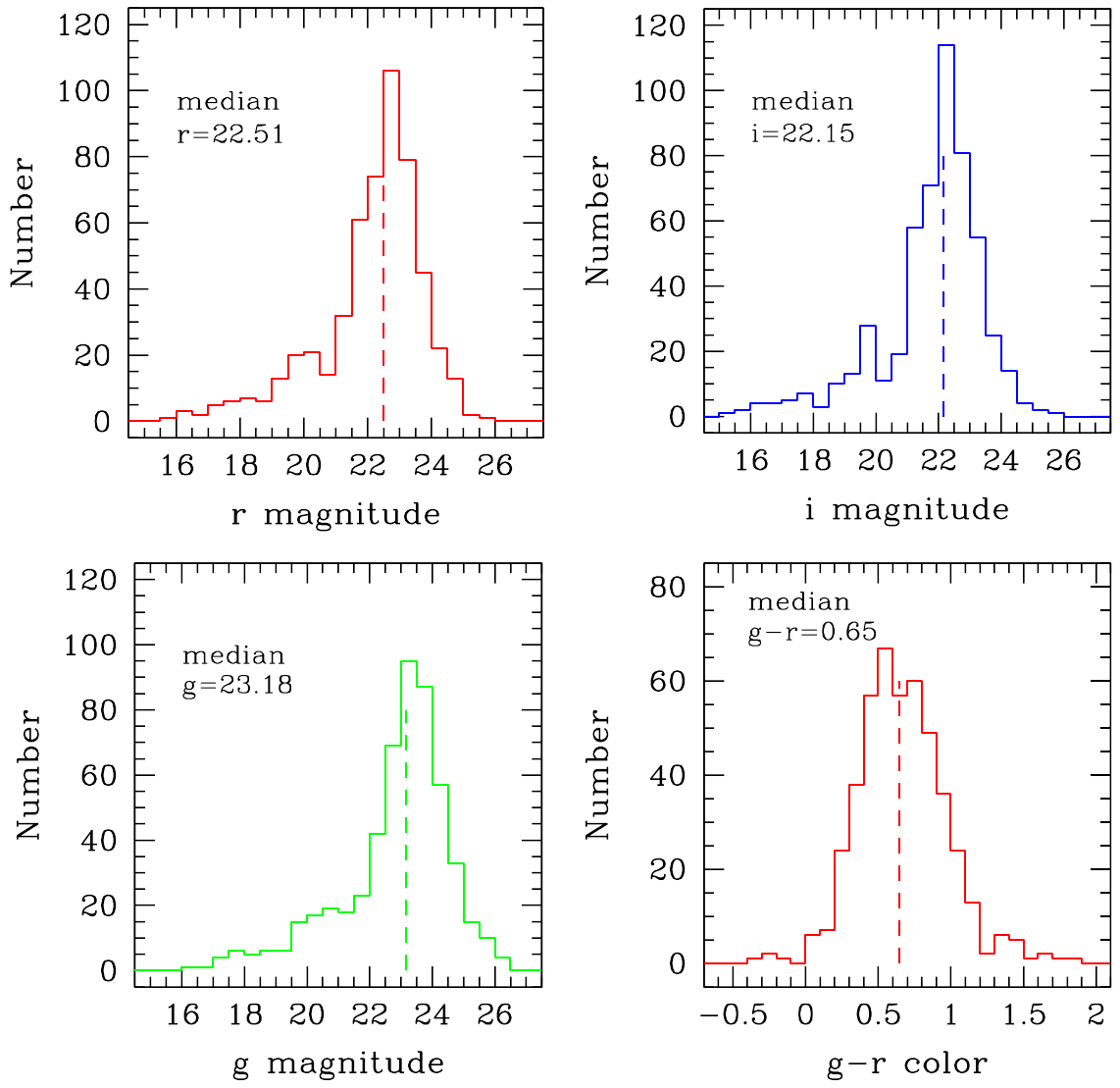}
\caption{Histograms showing the BB magnitudes and colors for the full sample of SFACT emission-line candidates from the three pilot-study fields (N = 533).   {\it upper left:} r-band magnitude distribution, with a median value of r = 22.51 and a faint limit of r $\sim$ 25.8.  {\it upper right:} i-band magnitude distribution, with a median value of i = 22.15.  {\it lower left:} g-band magnitude distribution, with a median value of r = 23.18 and a faint limit of g $\sim$ 26.3.   {\it lower right:} g$-$r color distribution, with a median value of 0.65.  While the bulk of the galaxies have observed colors between 0.2 $<$ g$-$r $<$ 1.2, many objects with extreme colors are present in the sample.}
\label{fig:phothists}
\end{figure}

Figure~\ref{fig:phothists} presents the results of our BB photometry for the 533 SFACT candidates in the pilot-study fields.  The details of our measurement and calibration procedures are given in SFACT2.    Here we show the cumulative histograms of the three BB magnitudes measured by the survey ({\it gri}), as well as the {\it g}$-${\it r} color distribution.   The median apparent magnitudes are listed in each panel of the figure, and the vertical dashed lines indicate the location of the median in each histogram.  

The apparent magnitude distributions shown in Figure~\ref{fig:phothists} reflect the depth of our sample.  Focusing on the r-band histogram (upper left), it is seen that the majority of SFACT ELG candidates have r-band magnitudes in the range 21-24 mag, with a tail to both brighter and fainter magnitudes.  The faintest object detected has r = 25.85, while the brightest has r = 15.60.  Nearly all of the objects with magnitudes brighter than r = 19.0 represent low redshift extended galaxies detected via their disk \ion{H}{2} regions (referred to as category 2 objects above).  The median value of r = 22.51 is 
consistent with {\it faintest} objects that are reliably detected by SDSS.
The g-band and i-band magnitude distributions are similarly deep, with median values of g = 23.18 and i = 22.15.

The histogram of {\it g}$-${\it r} colors shown in  Figure~\ref{fig:phothists} (lower right) exhibits a surprisingly broad range.  The median color of 0.65 is consistent with the colors of early-type spirals, and the color range that encompasses the bulk of the sample, 0.2 $<$ {\it g}$-${\it r} $<$ 1.2, includes many very red systems.  Typically one would expect that a sample of galaxies that is dominated by actively star-forming systems would exhibit bluer colors.  Two factors appear to be creating this counter-intuitive result.  First, the colors presented here are {\it observed} values; we have not applied any Galactic reddening corrections (expected to be small for these survey fields) or redshift-dependent corrections (K corrections).  Since the galaxy sample includes many systems at redshifts of up to 1.0, one should expect that the relevant K corrections will be significant.   Second, for a sample of galaxies detected via their strong emission lines, it would not be surprising if the emission lines themselves impacted the observed colors in a redshift-dependent fashion.  For example,  the presence of strong [\ion{O}{3}]$\lambda$5007 emission in the SDSS r filter is what leads to the detection of Green Pea galaxies using SDSS BB photometry \citep{gp}.  For the SFACT galaxies with spectroscopic follow-up, the [\ion{O}{3}]-selected galaxies (n = 179) have a median {\it g}$-${\it r} color of 0.82, compared with {\it g}$-${\it r} = 0.51 for the lower-redshift H$\alpha$-selected subset (n = 107).  Hence, we conclude that the presence of strong  [\ion{O}{3}] emission is skewing the observed {\it g}$-${\it r} colors redward for a significant fraction of the SFACT galaxies.

\subsubsection{Emission-Line Fluxes of the Candidates}

\begin{figure}
\centering
\includegraphics[width=3.35in]{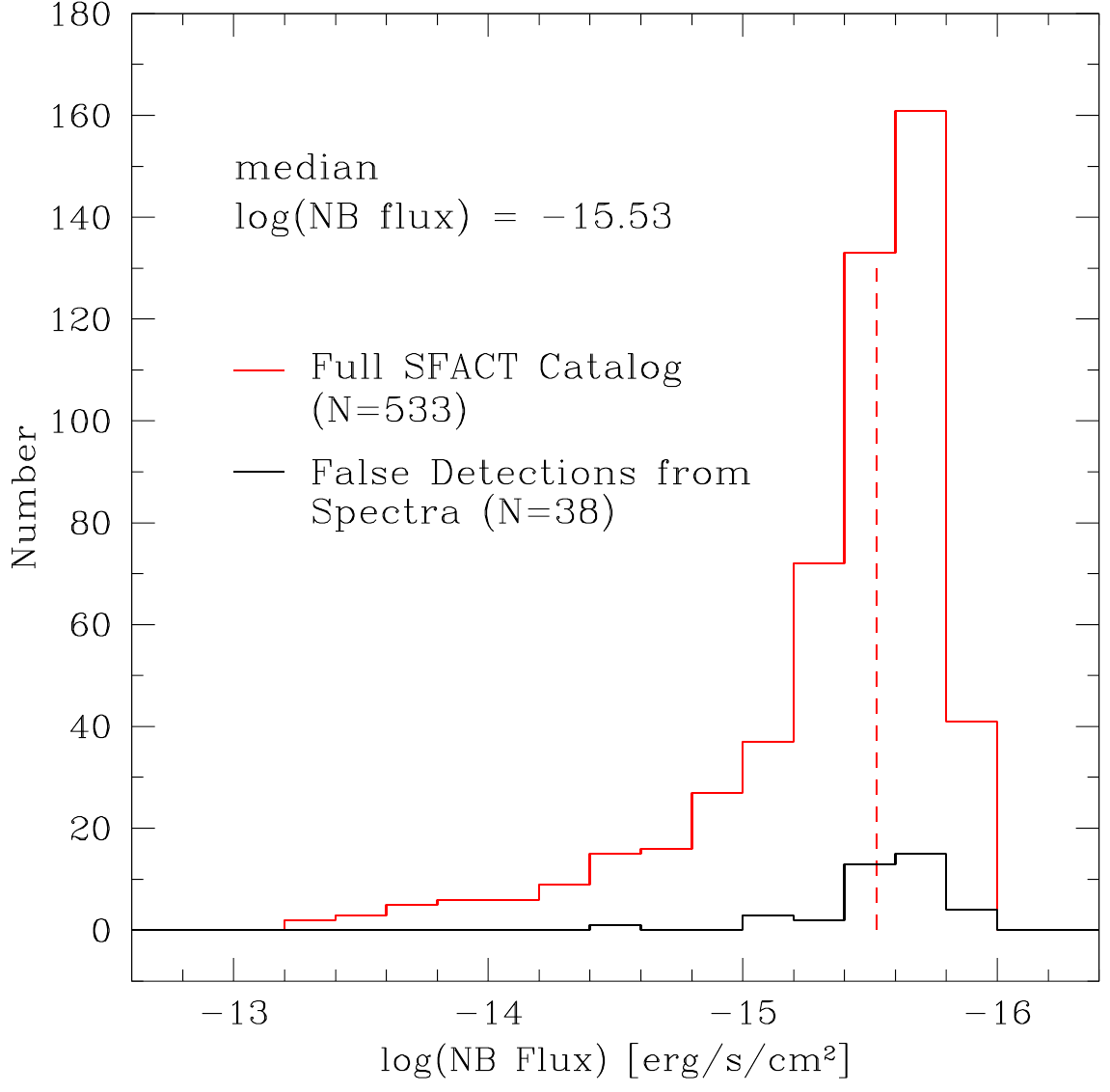}
\caption{Histogram of the NB fluxes measured for all 533 SFACT candidates from the three pilot-study fields.  We make no distinction between objects detected in the different NB filters.  The median emission-line flux (denoted by the dashed vertical line) is 2.97 $\times$ 10$^{-16}$ erg s$^{-1}$ cm$^{-2}$, while the sample appears to be substantially complete to $\sim$1.9 $\times$ 10$^{-16}$ erg s$^{-1}$ cm$^{-2}$.  The lower (black) histogram shows the distribution of NB fluxes for the 38 SFACT candidates that were found to be false detections in our follow-up spectroscopy (see Section 4.2)}
\label{fig:nbhist}
\end{figure}

While the BB magnitudes are a relevant and convenient way to parameterize the depth of the survey, SFACT is inherently an emission-line selected sample of galaxies.  Therefore, it is the emission-line flux distribution that truly defines the sensitivity limits of the survey.  Figure~\ref{fig:nbhist} plots the measured NB emission-line fluxes for all 533 SFACT objects in the pilot-study fields.   Once again, we refer the reader to the SFACT2 companion paper for details on the measurement and calibration procedures used to obtain the flux values shown in Figure~\ref{fig:nbhist}.

The histogram of emission-line fluxes reveals a strongly peaked distribution, with most sources exhibiting fluxes between 10$^{-15}$ and 10$^{-16}$ erg s$^{-1}$ cm$^{-2}$.   The median flux value is 2.97 $\times$ 10$^{-16}$ erg s$^{-1}$ cm$^{-2}$, while the minimum detected flux is 1.01 $\times$ 10$^{-16}$ erg s$^{-1}$ cm$^{-2}$.  The histogram of fluxes rises to a peak at $\sim$1.9 $\times$ 10$^{-16}$ erg s$^{-1}$ cm$^{-2}$, after which the number of detections begins to fall off.  We take this latter quantity to be representative of the completeness limit of the survey.

At the bright end of the distribution, all of the objects with fluxes above 10$^{-14}$ erg s$^{-1}$ cm$^{-2}$ (n = 16) are bright extended galaxies.   They are all nearby, detected via their H$\alpha$ emission in either NB1 or NB3.  In most cases, the line emission from these objects is a combination of disk \ion{H}{2} regions and nuclear emission from a central starburst.  The SFACT objects in the next decade of line flux (10$^{-14}$ to 10$^{-15}$ erg s$^{-1}$ cm$^{-2}$) represent a mixture of sources.  Of the 73 objects in this flux range, 30 are H$\alpha$-detected extended galaxies, similar in nature to the very brightest sources.   However, the remaining 43 include many compact [\ion{O}{3}]-detected objects, plus one luminous [\ion{O}{2}]-selected source and two bright QSOs.

\vfill\eject
\subsubsection{Example SFACT Objects}

\begin{figure*}
\centering
\includegraphics[width=6.5in]{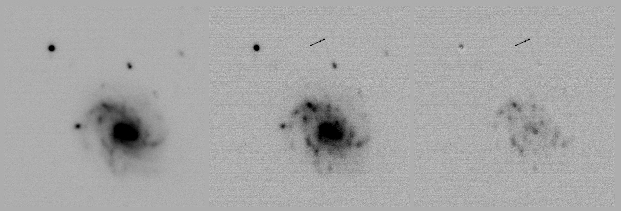}
\vskip -1.24in \hskip -4.30in {\huge \textcolor{red}{$\bigcirc$}}
\vskip -0.28in \hskip -0.02in {\huge \textcolor{red}{$\bigcirc$}}
\vskip -0.28in \hskip 4.28in {\huge \textcolor{red}{$\bigcirc$}}

\vskip 0.04in \hskip -3.90in {\huge \textcolor{red}{$\bigcirc$}}
\vskip -0.28in \hskip 0.40in {\huge \textcolor{red}{$\bigcirc$}}
\vskip -0.28in \hskip 4.70in {\huge \textcolor{red}{$\bigcirc$}}

\vskip 0.71in
\includegraphics[width=6.5in]{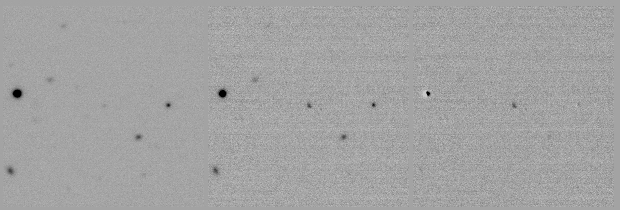} 
\vskip -1.24in \hskip -4.30in {\huge \textcolor{red}{$\bigcirc$}}
\vskip -0.26in \hskip -0.02in {\huge \textcolor{red}{$\bigcirc$}}
\vskip -0.27in \hskip 4.28in {\huge \textcolor{red}{$\bigcirc$}}
\vskip 0.97in

\includegraphics[width=6.5in]{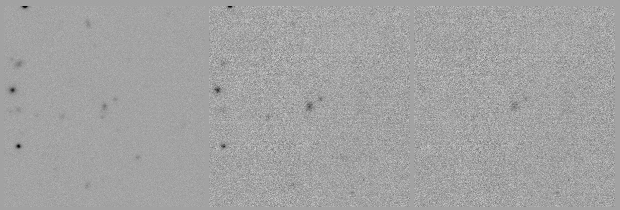} 
\vskip -1.23in \hskip -4.30in {\huge \textcolor{red}{$\bigcirc$}}
\vskip -0.26in \hskip -0.02in {\huge \textcolor{red}{$\bigcirc$}}
\vskip -0.27in \hskip 4.28in {\huge \textcolor{red}{$\bigcirc$}}
\vskip 1.00in

\caption{Example emission-line objects from the SFACT survey.   Each row shows a trio of images: left is the BB continuum image, center is the NB image before continuum subtraction, and right is the continuum-subtracted NB image.  Each image cutout shows a field-of-view of 50 $\times$ 50 arcsec.  {\it Top:}\ The spiral galaxy SFF10-NB1-C22247, which is marked by the lower set of red circles.  Also indicated is the \ion{H}{2} region SFF10-NB1-C22168.  {\it Middle:}\  The faint (g = 24.98) source SFF15-NB2-D2938.  Note how the object is dramatically brighter in the NB images.   {\it Bottom:}\  The ELG SFF10-NB1-D11121.  Spectra of all four SFACT sources illustrated in this figure are presented in Figure~\ref{fig:spec1}}
\label{fig:images1}
\end{figure*}

Figure~\ref{fig:images1} presents examples of newly discovered ELGs found by SFACT in the pilot-study fields.   We present these images to illustrate the survey method as well as to give an indication of the data quality.   Each row in the figure presents three 50 $\times$ 50 arcsec cutouts.  The left image in each row shows the BB image of the field that was used in the continuum subtraction.  In all cases shown, this would be the sum of the r-band and i-band images.  The middle figure is that of the NB image within which the emission-line object was detected, shown before any continuum subtraction.   The rightmost figure is that of the continuum-subtracted NB image.   The red circles displayed in each panel identify the SFACT objects.

The top row of images illustrate an extended galaxy as detected by SFACT.   The galaxy shown is SFF10-NB1-C22247; this designation refers to the center of the galaxy.   This particular object was detected via H$\alpha$ emission in the NB1 filter.  It has a total g-band magnitude of 16.59 and an integrated NB flux of 2.14 $\times$ 10$^{-14}$ erg s$^{-1}$ cm$^{-2}$.   These characteristics make SFF10-NB1-C22247 one of the very brightest sources cataloged in the three pilot-study fields in both BB and NB flux.   As described in  \S 4.1.2, we often include a bright \ion{H}{2} region located in the disk of an extended galaxy such as this in the catalog.  We do this in cases where the \ion{H}{2} region exhibits substantial emission and/or in cases where the emission from the galaxy center is weak or nonexistent.  In the case shown here, the \ion{H}{2} region (designated SFF10-NB1-C22168) is the brightest knot of emission within the galaxy.

The second row of Figure~\ref{fig:images1} presents the survey images for one of the faintest sources in the pilot study.  This galaxy, SFF15-NB2-D2938, is barely visible in the continuum image, but is readily evident in the NB image.  This source, which was detected in NB2, has a g-band magnitude of only 24.98 and a NB flux of 3.26 $\times$ 10$^{-16}$ erg s$^{-1}$ cm$^{-2}$.  This object provides a wonderful illustration of one of the aspects of our data analysis process: due to the manner with which we carry out our searches for emission-line objects, SFACT is capable of detecting ELGs with little or no continuum emission (see SFACT2).

The final object shown in Figure~\ref{fig:images1} is  less extreme in its appearance than the other two, making it perhaps more representative of a typical SFACT detection.  This source, SFF10-NB1-D11121, was detected as an emission-line object in filter NB1.  Its observed properties include a g-band magnitude of 23.28 and a NB flux of 4.76 $\times$ 10$^{-16}$ erg s$^{-1}$ cm$^{-2}$.   The line emission is clearly spatially extended.  While the emission is strong,  it does not exhibit the high equivalent width evident in SFF15-NB2-D2938.

The spectra for all of the SFACT detections shown here are presented in Figure~\ref{fig:spec1}.   Many additional example images of SFACT ELGs are given in SFACT2.

\subsection{Results from the Spectroscopic Follow-Up Observations}

Once an SFACT field has been processed and searched for ELG candidates, it must still be observed spectroscopically in order to make it useful for science.  Because SFACT is sensitive to faint sources via a variety of emission lines, spectroscopic confirmation is required to ascertain which line a given object was detected with.  Lower redshift objects are often, but not always, sufficiently resolved in our images to allow us to identify them as H$\alpha$-detections.  However, for most of the higher redshift sources, including [\ion{O}{3}] and [\ion{O}{2}] detections, the sources are unresolved or only marginally resolved in our images.  Hence, we typically cannot identify the emission line present in our survey filter without the aid of spectra.  In addition, spectroscopy is required to identify the activity class of each source (e.g., star forming {\it vs.} AGN).

As alluded to in \S 3.1, follow-up spectroscopy of the SFACT candidates was planned for from the start of the project.  The predicted density of ELG candidates as well as the footprint of the ODI camera was well matched for the Hydra fiber positioner available on the WIYN 3.5 m telescope.  Despite the faint nature of the SFACT objects, the fact that their spectra are dominated by emission lines makes it possible to acquire adequate spectra for even the faintest sources.  Details of the spectroscopic observations, data processing, and measurement are presented in a companion paper (SFACT3).   Here we present a brief overview of the essential results of the spectroscopic observations of the three pilot-study fields.

\subsubsection{Number of Spectroscopically Observed SFACT Objects}

Naturally, the follow-up spectroscopy of our candidates lags the imaging observations.   The earliest opportunity to acquire spectra of the SFACT candidates occurs one year after the full set of imaging data for a given field have been obtained.  However, in practice it has usually taken substantially longer to get complete spectroscopic data for all of the candidates.   The reasons for this are varied (e.g., loss of telescope access for two semesters due to the Covid-19 pandemic, weather losses, etc.).   Hence, the follow-up spectroscopy of the three pilot-study fields is not yet complete.  Nonetheless, it is sufficiently far along to allow us to present a reasonably robust picture of the nature of the SFACT objects.

Table~\ref{tab:pilot_study} includes the number of objects for which follow-up spectra have been obtained.   Overall, 453 out of 533 objects have been observed spectroscopically (85.0\%).  Each of the individual fields has a substantial fraction of its ELG candidates observed to date: 83.3\% for SFF01, 93.1\% for SFF10, and 78.9\% for SFF15.

For the SFACT objects with spectral observations in our pilot-study fields, 415 out of 453 (91.6\%) are found to be true emission-line objects.  That is, 91.6\% of our sources exhibited moderate-to-strong line emission located within the NB filter where the excess flux signal was present in our survey images.   The 8.4\% false-detection rate is in keeping with expectations, given that we chose to be inclusive in terms of the quality of the sources that were passed during our final object inspection (see SFACT2).  In most cases, the false detections represent faint sources with weaker putative emission that exhibit properties that locate them near the thresholds of our selection criteria (see Figure~\ref{fig:nbhist}).

It is worth noting that only a small fraction of the SFACT objects have existing spectra in SDSS.   This is expected, given that the SDSS galaxy redshift survey only observed galaxies brighter than  r $\sim$ 18 \citep{strauss2002}, while only a small fraction of the SFACT galaxies are brighter than r = 18 (see Figure~\ref{fig:phothists}).  Since the SDSS galaxy redshift survey was not carried out in the Fall sky, only 2 out of 533 SFACT candidates found in the pilot-study fields also possess SDSS spectra (0.4\%).   For 13 Spring SFACT fields where the catalog construction is complete, 61 out of 1659 objects (3.7\%) have spectra in SDSS.   Of these, 14 are QSOs, while 47 are galaxies.   Among the latter group, 100\% are detected by SFACT via their H$\alpha$ line (redshifts below z = 0.15), and nearly all are brighter than r = 18.

\subsubsection{Example Spectra}

\begin{figure*}
\centering
\includegraphics[width=3.35in]{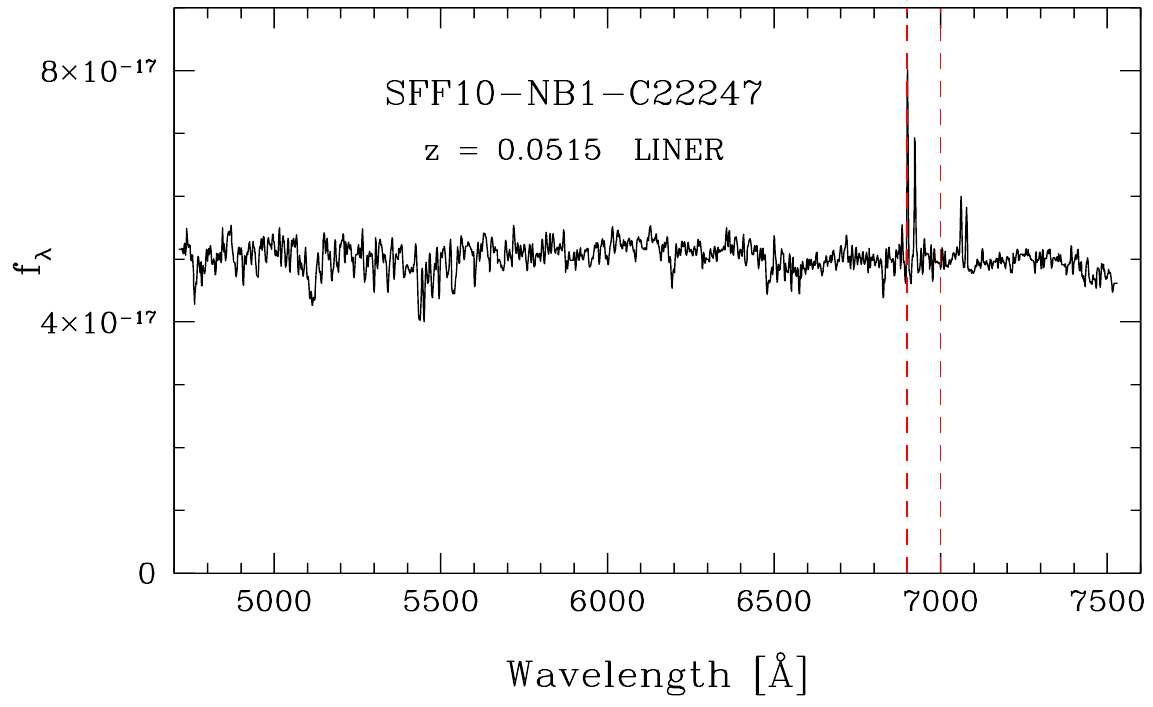} \includegraphics[width=3.35in]{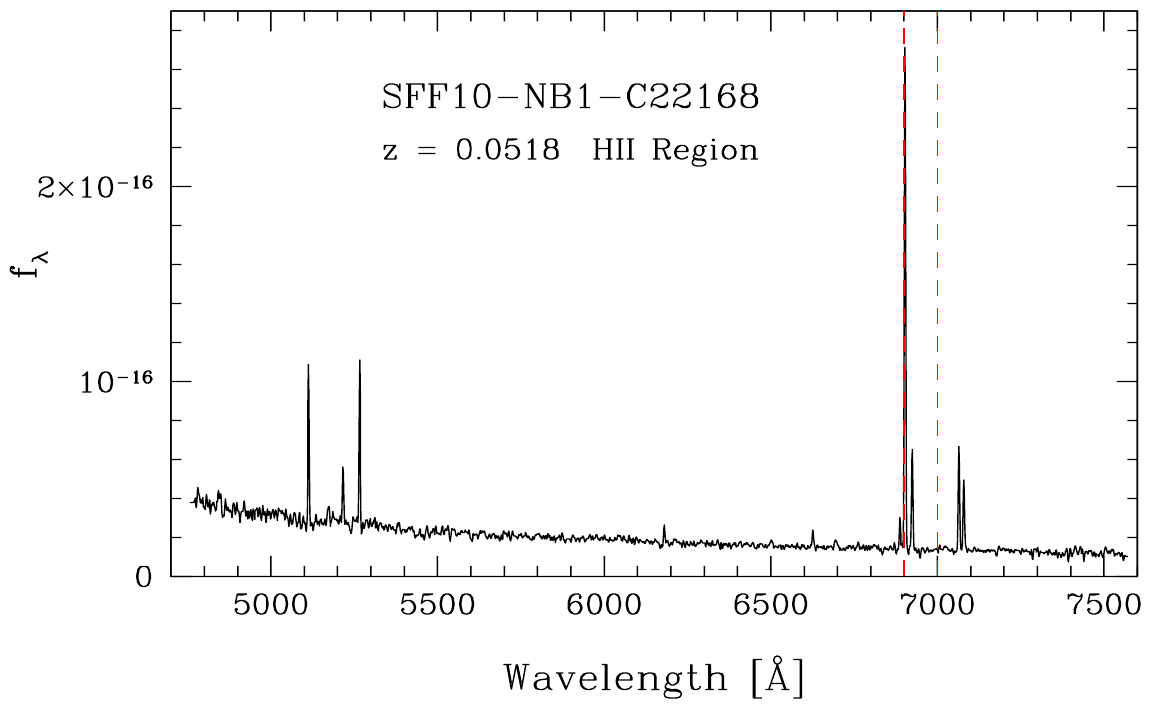}
\includegraphics[width=3.35in]{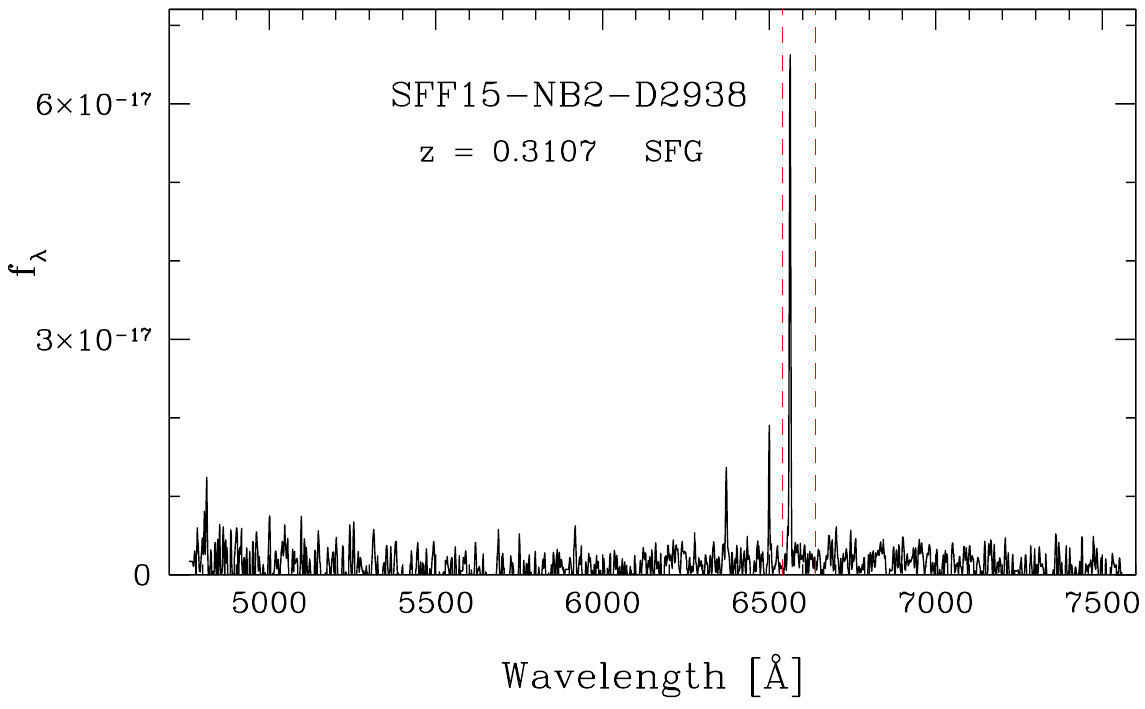}   \includegraphics[width=3.35in]{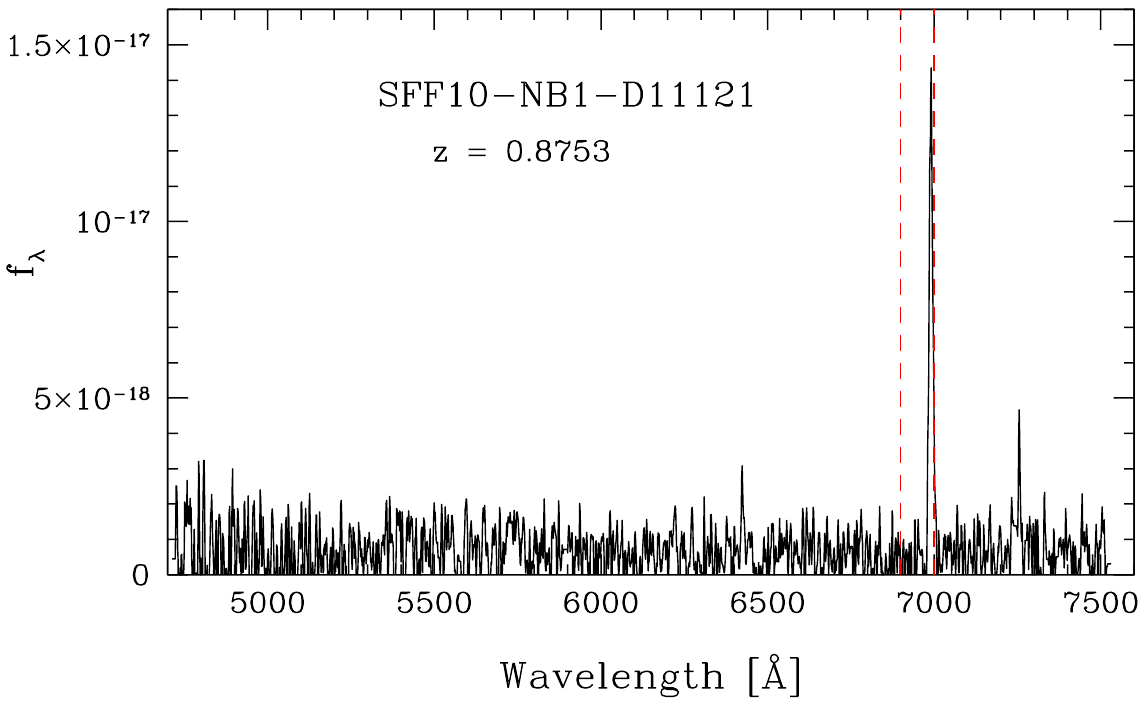} 
\caption{Spectral plots of the objects illustrated in Figure~\ref{fig:images1}.  The y-axis plots the flux in erg s$^{-1}$ cm$^{-2}$ \AA$^{-1}$, while the x-axis shows the wavelength in \AA.  The vertical red dashed lines found in all spectral plots presented in this paper indicate the FWHM bandwidth of the NB filter used to detect each object whose spectrum is shown.  The upper row shows the spectra of both the central region (left) and an outlying \ion{H}{2} region (right) for the spiral galaxy shown in the top image of Figure~\ref{fig:images1}.  The nuclear spectrum indicates that this object is a LINER.   The lower row includes the spectrum of an [\ion{O}{3}]-detected star-forming galaxy found in the NB2 filter (left), and a higher-redshift [\ion{O}{2}]-detected system (right). }
\label{fig:spec1}
\end{figure*}

Figure~\ref{fig:spec1} presents example spectra of objects discovered in our survey.  The spectra shown correspond to the SFACT candidates illustrated in Figure~\ref{fig:images1}.  Our SFACT spectra typically cover the wavelength range between $\sim$4700 and 7600 \AA.  The objects whose images and spectra are shown in Figures~\ref{fig:images1} and~\ref{fig:spec1} were chosen primarily to illustrate SFACT ELGs detected by the three principal emission lines selected for in the survey: H$\alpha$, [\ion{O}{3}]$\lambda$5007, and [\ion{O}{2}]$\lambda$3727.  

The top row of Figure~\ref{fig:spec1} shows the spectrum of the central region of the spiral galaxy shown in Figure~\ref{fig:images1}, as well as a spectrum of one of the outlying  \ion{H}{2} regions detected in our imaging data.  As indicated in \S 4.1.2, the SFACT survey always includes the center of any extended galaxy for which disk \ion{H}{2} regions are identified in our imaging data.   We also often include the brightest \ion{H}{2} region in our catalogs, particularly if the central region has only weak or nonexistent line emission indicated in our NB images.   In this case, the central region of the galaxy includes a moderately bright, high-surface-brightness nucleus which harbors a Low-Ionization Nuclear Emission Region \citep[LINER;][]{heckman1980}.  The spectrum of the  \ion{H}{2} region reveals a normal, relatively metal-rich star-forming knot.  The galaxy has a redshift of z = 0.0515 (distance = 232 Mpc), and has an absolute magnitude of M$_g$ = $-$20.2.

The [\ion{O}{3}]-detected object whose spectrum is shown in the lower left portion of Figure~\ref{fig:spec1} is a low-luminosity star-forming galaxy with a redshift z = 0.3107.  The strong [\ion{O}{3}]$\lambda$5007 emission line was detected in the NB2 filter.  We measure a g-band magnitude of 24.98, from which we derive an absolute magnitude of M$_g$ = $-$16.1.  It is seen to be essentially unresolved in the survey imaging data shown in Figure~\ref{fig:images1}.  This object has physical and spectral properties that are reminiscent of blue compact dwarfs (BCDs) in the local universe \citep[e.g.,][]{jano2014, jano2017}.  However, SFACT has detected it at a distance of over 1600 Mpc.

The final example spectrum presented in Figure~\ref{fig:spec1} is that of an [\ion{O}{2}]-detected object (lower right).   As is clear from the figure, the spectra of the [\ion{O}{2}]-selected SFACT objects typically do not provide access to many of the nebular emission lines that both aid in the line identification process and provide vital diagnostics regarding the nature of the source.  We sometimes detect the [\ion{Ne}{3}]$\lambda$3869 line, as is the case with the object shown in Figure~\ref{fig:spec1}.  But in many cases we only detect the [\ion{O}{2}] doublet itself.  The identification of the detected emission line is seldom in doubt, however, because the doublet is typically marginally resolved at our spectral resolution.  Due to the lack of meaningful emission-line diagnostics, we typically cannot assign the [\ion{O}{2}]-detected ELGs into an activity class.  This step will need to await the acquisition of longer-wavelength spectral data that cover the rest-frame spectra up to $\sim$5000 \AA.  Hence, at the current time, we can only partially assess the nature of this object.   With a redshift of z = 0.8753 and an absolute magnitude M$_g$ = $-$20.5, we know that this galaxy is quite luminous.   Our guess in that this is a starburst galaxy, as opposed to an AGN, based on the high [\ion{O}{2}]/[\ion{Ne}{3}] flux ratio, the narrowness of the emission lines, and the fact that the [\ion{Ne}{5}]$\lambda\lambda$3426,3346 lines are not present. 

Additional example images and spectra of both SFACT galaxies and QSOs can be found in SFACT2 and SFACT3.

\subsubsection{Redshift Distribution}

\begin{figure}
\centering
\includegraphics[width=3.35in]{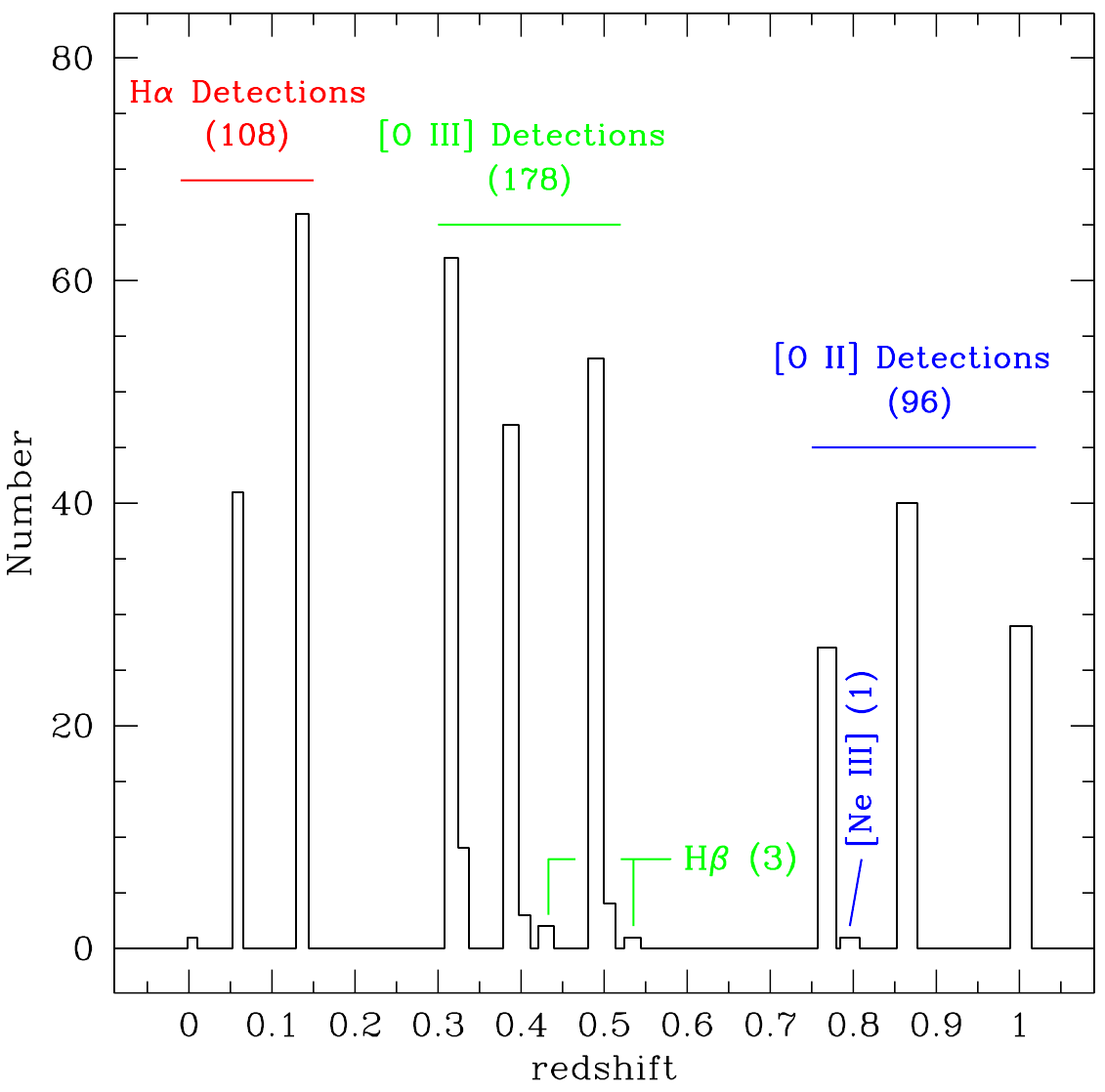}
\caption{Histogram showing the redshift distribution of the SFACT ELGs.   Only galaxies detected via their H$\alpha$, [\ion{O}{3}], H$\beta$, [\ion{O}{2}] and [\ion{Ne}{3}] lines are included in the figure; higher redshift QSOs are excluded.  Unlike a typical magnitude-limited redshift survey, the SFACT survey samples the universe in discrete redshift windows}
\label{fig:zhist}
\end{figure}

The redshifts of the SFACT galaxies detected in our pilot-study fields are shown in Figure~\ref{fig:zhist}, presented in quasi-histogram form.  Only objects detected via one of the strong optical nebular lines 
are included; higher redshift QSOs are not shown in this figure.  The plot is designed to  graphically illustrate the redshift windows sampled by the survey filters for the principal emission lines.  We note that the width of each bin corresponds to the width of each redshift window, getting broader with increasing redshift.   Specifically, these widths correspond to the redshift range of objects contained within the half-height width of the relevant filter (see Table~\ref{tab:filters}).  

The redshift windows shown in the histogram break down into three primary groups.  The lowest-redshift objects (0.0 $<$ z $<$ 0.15) are the H$\alpha$-detected galaxies.  Many of these represent large, extended galaxies with multiple \ion{H}{2} regions in their disks.  As suggested earlier, we expect the number of H$\alpha$ detections in NB2 to be extremely small.  Only one such object exists in the three pilot-study fields: SFF01-NB2-B19198 is an \ion{H}{2} region in a nearby dwarf irregular galaxy (z = 0.0034).  The other two NB filters have a significant number of H$\alpha$ detections.  In particular, the NB3 H$\alpha$ detections represent the single largest sample of ELGs in the current sample (n = 65).  

The next grouping in Figure~\ref{fig:zhist} exhibits a more complex structure in the histogram because the galaxies are being detected via three different lines.  The tallest bins represent the galaxies detected via their [\ion{O}{3}]$\lambda$5007 line.  Unlike the situation for the H$\alpha$-detected ELGs, the effective volumes covered by the three NB filters for the [\ion{O}{3}] line do not differ by large amounts.  Hence, the numbers of objects detected in the three filters should be comparable (although modulated somewhat by cosmic variance).  This is in fact what we observe.  The number of [\ion{O}{3}]$\lambda$5007 detections are NB2 = 62, NB1 = 47, and NB3 = 53.   The total number of [\ion{O}{3}]$\lambda$5007-detected galaxies (n = 162) represents the largest number for a specific line in the sample.

The second set of galaxies in this region of the histogram represent objects that are detected via the [\ion{O}{3}]$\lambda$4959 line.  These are galaxies where the $\lambda$5007 line has redshifted out of the NB filter, but the $\lambda$4959 line is still present.  Since the latter line is roughly a factor of 2.9$\times$ weaker than the former \citep[e.g.,][]{agn2}, we naturally expect far fewer detections via this line.  This is what is observed: a total of 16 ELGs are detected via their [\ion{O}{3}]$\lambda$4959 line, only $\sim$10\% the number of [\ion{O}{3}]$\lambda$5007 detections.   Finally, there are three galaxies that are detected by their H$\beta$ emission (two in NB1 and one in NB3).  The  H$\beta$-detected objects represent only a small contribution to the overall survey.   

When the need to distinguish between which of the three lines was detected in the survey is of less relevance, the objects in this grouping will be lumped together and simply referred to as being [\ion{O}{3}]-detected.  However, for many applications the distinction as to the proper identification of the detected line is essential (e.g., calculating star-formation rate densities).  Whenever relevant, we will separate the galaxies detected by these three lines into distinct subsamples.

The final grouping in Figure~\ref{fig:zhist} represents the highest redshift group, nearly all of which are detected via the [\ion{O}{2}]$\lambda$3727 doublet.  ELGs with redshifts in the range 0.75 to 1.02 are included within this group.  At these distances, all of the detected objects are unresolved in our images.   Naturally, the [\ion{O}{2}]-detected component of the survey are predominantly among the faintest objects in the sample.  As was the case with the [\ion{O}{3}]-detected component of the survey, the [\ion{O}{2}] ELGs are fairly evenly distributed between the three filters: NB2 = 26, NB1 = 42, and NB3 = 28.  While the total number of [\ion{O}{2}] detections (n = 96) is lower than both the H$\alpha$ and [\ion{O}{3}] subsamples, it nonetheless represents a sizable fraction of the total number of SFACT objects ($\sim$25\% of the ELGs included in Figure~\ref{fig:zhist}).   We also note that a single object was detected due to the presence of a strong [\ion{Ne}{3}]$\lambda$3869 line falling in the NB1 filter.

There are 11 spectroscopically confirmed QSOs detected in the SFACT pilot-study fields.  Of these, one was detected via the [\ion{O}{2}] doublet (z = 0.756), five were selected when \ion{Mg}{2} $\lambda$2798 fell in one of the survey filters (redshifts between 1.34 and 1.68), two were found through their \ion{C}{3}] $\lambda$1908 emission (redshifts between 2.43 and 2.94), and three were detected via the \ion{C}{4} $\lambda$1549 line (redshifts between 3.23 and 3.85).    The number of detected QSOs is small compared to the number of ELGs discovered by SFACT, but this is to be expected given the relatively small volumes surveyed by our three pilot-study fields combined with the small volume densities of QSOs.  We project that we will detect over 200 QSOs in the full survey.

\subsection{Properties of the SFACT Emission-Line Objects: A Preliminary View}

Despite the incomplete nature of the spectroscopic follow-up of the SFACT galaxies detected in our pilot-study fields, we can nonetheless obtain a fairly good understanding of the nature of the sources detected in our sample.   In this section we present a preliminary look at the properties of the SFACT galaxies.   A more complete assessment of the nature of the sources detected in our survey will be presented in future papers in this series.

\subsubsection{Absolute Magnitude Distributions}

\begin{figure}
\centering
\includegraphics[width=3.38in]{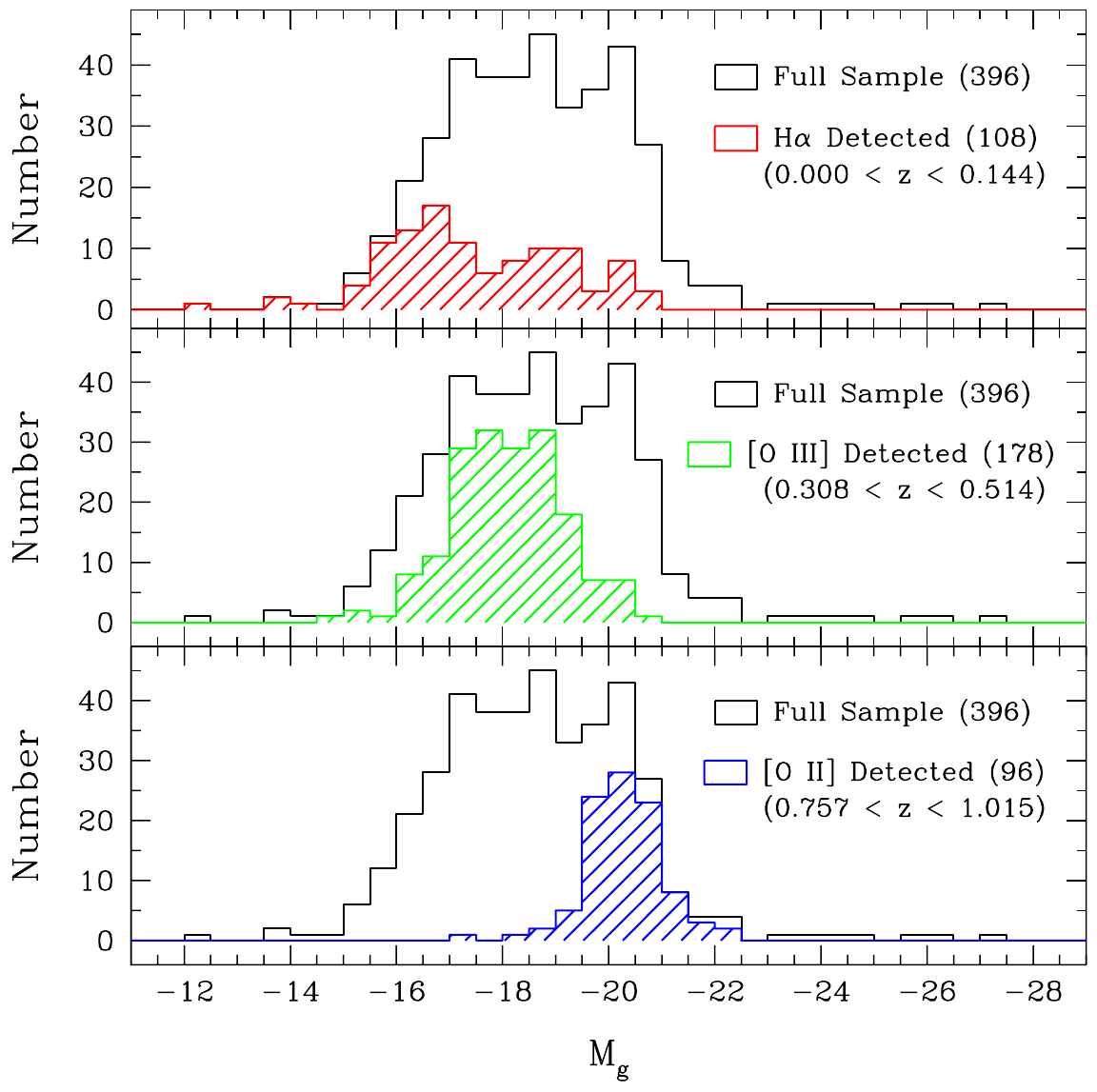}
\caption{Histograms showing the g-band absolute magnitude distributions for the SFACT objects located in our pilot-study fields. The upper panel shows the M$_g$ distribution for the lower-redshift H$\alpha$-detected galaxies, while the middle and lower panels show the same distributions for the intermediate-redshift [\ion{O}{3}]-selected galaxies and the [\ion{O}{2}]-selected galaxies, respectively.   In all three panels the black-lined histogram plots the luminosity distribution for the full sample.   The latter includes  the higher-luminosity QSOs. }
\label{fig:absmaghist}
\end{figure}

In Figure~\ref{fig:absmaghist} we present histograms of the g-band absolute magnitudes of the SFACT galaxies with redshifts in the three pilot-study fields.  The galaxies are separated into three groups based on the emission line responsible for their detection.   Each panel of the figure also shows the luminosity distribution for the full sample (black-lined histogram).  
Objects detected as \ion{H}{2} regions are excluded from Figure~\ref{fig:absmaghist}.  Rather, we show the luminosities of the entire galaxy within which the \ion{H}{2} regions reside.  

The overall distribution of absolute magnitudes is remarkably broad, with the range $-$16 $>$ M$_g$ $>$ $-$21 being well represented and exhibiting a fairly flat array of values.   This is characteristic of emission-line-selected galaxy samples in general \citep[e.g.,][]{um2, hadot2}, since the detection of lower-luminosity ELGs is typically enhanced in such surveys.   This is in contrast to the luminosity distributions of traditional magnitude-limited galaxy surveys, which tend to be strongly peaked with the majority of galaxies falling within $\pm$1 magnitude of M$^*$ (for reference, M$^*_g$ = $-$20.1; \citet{blanton2003}).  We conclude that SFACT, like previous emission-line surveys, is quite sensitive to low-luminosity dwarf systems, particularly for the low- and intermediate-redshift detections.

The upper panel of Figure~\ref{fig:absmaghist} highlights the M$_g$ distribution of the H$\alpha$-detected SFACT galaxies.  These galaxies are found at lower redshifts (see Figure~\ref{fig:zhist}), and represent the most diverse subset of the overall survey.  Many of the galaxies are larger spirals or irregulars with multiple \ion{H}{2} regions detected in our images.  These represent the more luminous galaxies shown in the upper panel (M$_g$ $<$ $-$19).  The lower luminosity galaxies in the histogram are typically compact star-forming systems of the type that are commonly labeled as BCDs in the nearly universe.  With one exception (see below), the H$\alpha$-selected ELGs are all detected in NB1 (distances of 230--300 Mpc) and NB3 (distances of 615--685 Mpc).   It is a testament to the depth of the SFACT survey method that it can routinely detect dwarf star-forming galaxies with such low luminosities at these distances.   For example, the second-lowest luminosity galaxy in this subsample was detected in NB3.   It has an absolute magnitude of M$_g$ = -13.97 and a distance of 628 Mpc.

The single exception mentioned in the previous paragraph is SFF01-NB2-B19198, the one NB2 H$\alpha$-detection in the three pilot-study fields.  This galaxy is a fairly low-surface-brightness dwarf with a single \ion{H}{2} region.  Based on its redshift, it has a distance of 18.9 Mpc and a g-band absolute magnitude of $-$12.38.  This makes it the lowest luminosity system in the current study.  

The luminosity histogram of the [\ion{O}{3}]-selected SFACT galaxies is shown in the middle panel of Figure~\ref{fig:absmaghist}.  The distribution of values is very symmetric, with a median M$_g$ of $-$18.1.  This relatively low-luminosity median value (2 magnitudes below M$^*_g$) is perhaps surprising at first glance, given that the distances to the galaxies in this subsample range between 1610 and 2970 Mpc.  However, it is important to recognize that any [\ion{O}{3}]-selected galaxy sample will be subject to a well understood metallicity effect: the [\ion{O}{3}] emission lines are weak in high-metallicity, high-luminosity systems and become stronger in lower metallicity systems.  The strength of the [\ion{O}{3}] doublet peaks at metal abundances of $\sim$10\% solar, abundances typically found in galaxies with $-$16 $>$ M$_g$ $>$ $-$19.  This is precisely the luminosity range occupied by the bulk of the [\ion{O}{3}]-detected subsample.   Most of these systems will be intermediate- and low-luminosity star-forming galaxies, including some BCDs at values of M$_g$ $>$ $-$17.  There is a modest-sized tail of higher luminosity galaxies among the [\ion{O}{3}]-selected SFACT galaxies.  These include a few Seyfert 2 galaxies as well a number of putative Green Pea-like galaxies with intermediate and high luminosities and large [\ion{O}{3}] equivalent widths ($>$ 200 \AA).

The bottom panel in Figure~\ref{fig:absmaghist} highlights the [\ion{O}{2}]-selected galaxies.  Not surprisingly, these objects dominate the high luminosity end of the distribution.  In fact, unlike the case for the other two emission lines illustrated in the figure, the luminosity histogram of the [\ion{O}{2}]-detected SFACT objects is reminiscent of the distributions seen in magnitude-limited samples.  The median value of M$_g$ is $-$20.3.  The distribution of absolute magnitudes can be understood as a combination of the extreme distances involved for the [\ion{O}{2}] subsample (4.7 to 6.9 Gpc) plus the fact that the strength of the [\ion{O}{2}] doublet does not exhibit as strong a metallicity dependence as [\ion{O}{3}]$\lambda\lambda$5007,4959.  The observed distribution has tails to both low- and high-luminosities.   The high-luminosity systems most likely include several AGN.  At the present time our follow-up spectra do not include sufficient spectral-line diagnostics to cleanly distinguish between star-forming galaxies and AGN for the [\ion{O}{2}]-detected SFACT objects.  Survey plans include a second round of spectroscopy for the [\ion{O}{2}]-selected galaxies that will include wavelength coverage redward of our initial follow-up spectra (see \S 7).  On the low-luminosity side of the distribution, we observe that SFACT is sensitive to some fairly low-luminosity systems at these large distances, albeit in small numbers.   The lowest luminosity galaxy detected via the [\ion{O}{2}] doublet has M$_g$ = $-$17.5.

\subsubsection{Emission-Line Diagnostic Diagrams}

The ratios of strong emission lines have long been used as diagnostics for the physical conditions present in star-forming galaxies and AGN \citep[e.g.,][]{bpt, vo1987}.  In particular, commonly used diagnostic diagrams allow astronomers to distinguish between the two primary ionization sources present in ELGs (hot-star photo-ionization and black hole accretion disks) and provide estimates of the metal abundance of the hot gas.

The specific redshift ranges present in our catalog, combined with the choice of spectral coverage for our follow-up spectra, results in the need to use different diagnostic diagrams for the different emission lines responsible for the detection of the sources.  For the H$\alpha$-selected SFACT galaxies, our spectral wavelength coverage includes the [\ion{N}{2}]/H$\alpha$ and [\ion{O}{3}]/H$\beta$ ratios, but excludes the [\ion{O}{2}] doublet (as well as the [\ion{S}{2}]$\lambda\lambda$6731,6716 doublet for NB3 detections).  Hence for the H$\alpha$-detected sources we utilize what is commonly referred to as the BPT diagram \citep{bpt}: [\ion{O}{3}]/H$\beta$ {\it vs.} [\ion{N}{2}]/H$\alpha$.  For all of our [\ion{O}{3}]-selected SFACT galaxies, the lines in the vicinity of H$\alpha$ are redshifted beyond the high-wavelength end of our spectral coverage.   In this case, the only viable option is to use the  [\ion{O}{3}]/H$\beta$ {\it vs.} [\ion{O}{2}]/[\ion{O}{3}] diagram.   As alluded to above, our current spectral coverage does not provide for any useful diagnostics for the [\ion{O}{2}]-detected galaxies; further assessment of the nature of these galaxies will need to await additional spectral data that cover redder wavelengths (see \S 7).

\begin{figure}
\centering
\includegraphics[width=3.35in]{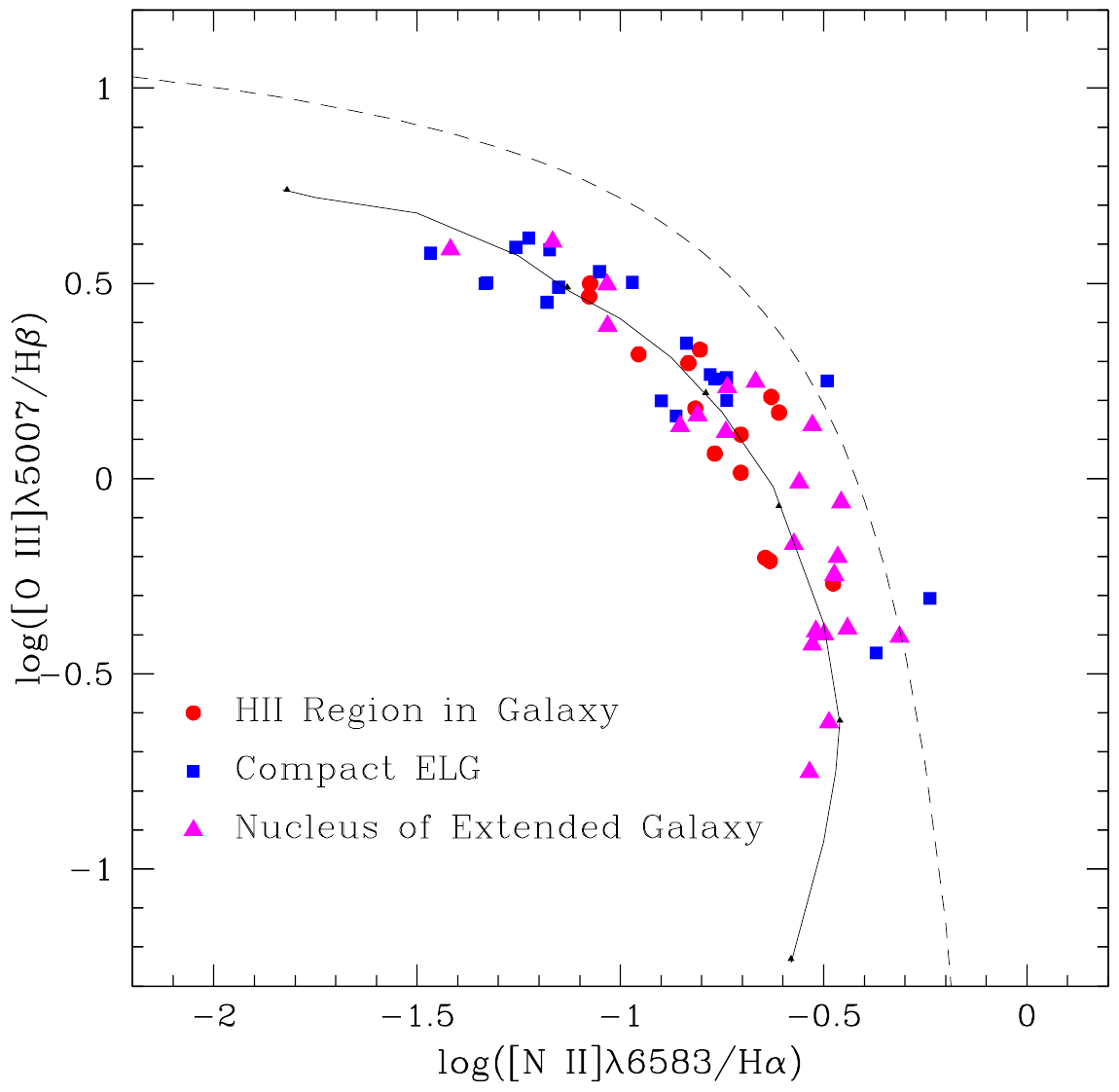}
\caption{Emission-line diagnostic diagram for the H$\alpha$-selected SFACT galaxies presented in the current paper.  These objects are found in the redshift range z = 0.00 -- 0.15.  The different categories of emission-line sources are indicated by different symbols, as specified in the legend.  The solid curve represents photo-ionization models from \citet{DE86}, and the dashed curve shows the empirical demarkation line between AGN and star-forming galaxies from \citet{kauff03}. }
\label{fig:diagplot1}
\end{figure}

\begin{figure}
\centering
\includegraphics[width=3.35in]{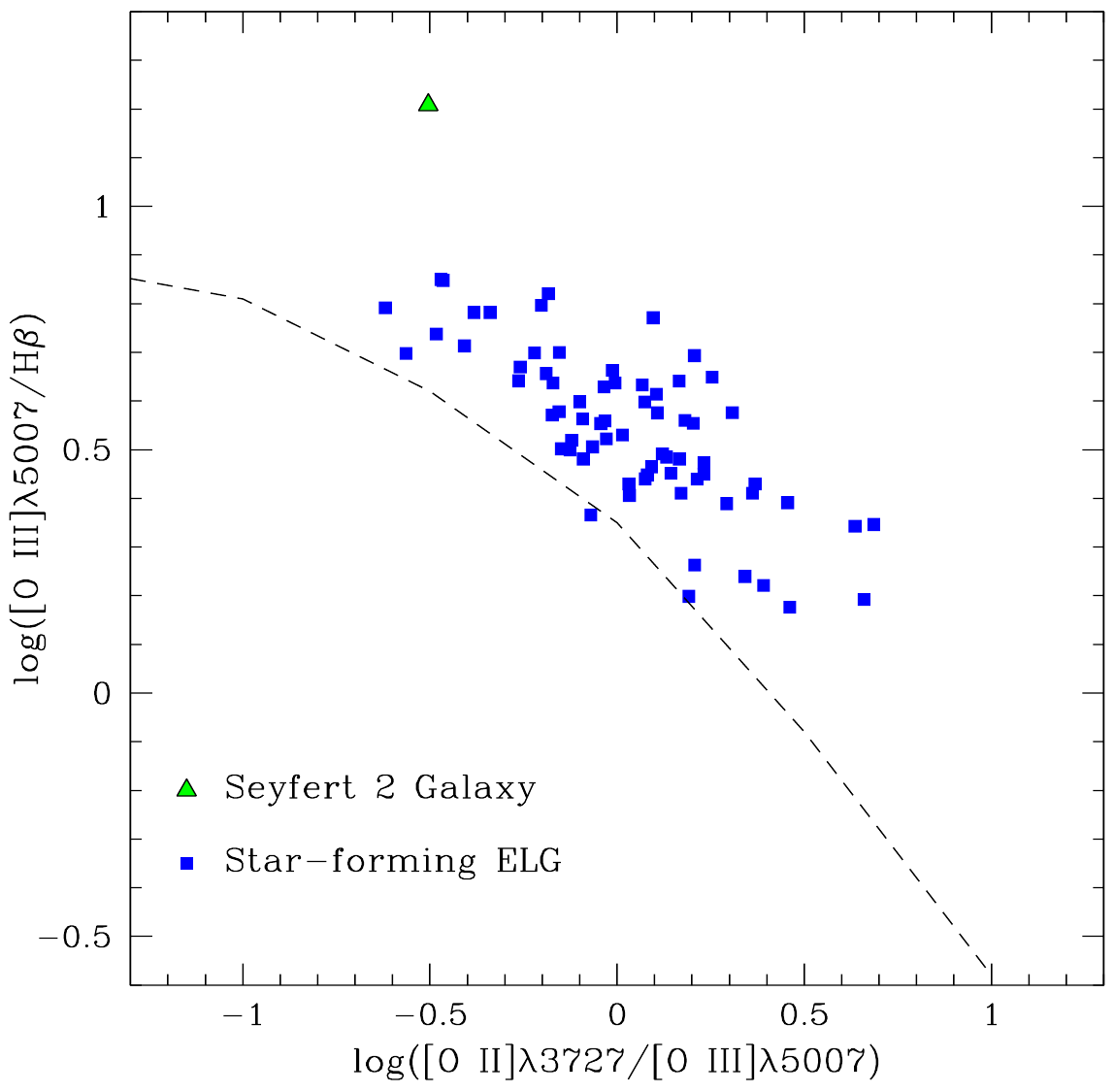}
\caption{Emission-line diagnostic diagram for the [\ion{O}{3}]-selected SFACT galaxies presented in the current paper.  These galaxies are found in the redshift range z = 0.30 -- 0.52.  The different types of emission-line sources are indicated by different symbols, as specified in the legend.  The dashed line represents the trend line shown in Figure 1 of \citet{bpt}.}
\label{fig:diagplot2}
\end{figure}

Figure~\ref{fig:diagplot1} shows the emission-line diagnostics for the H$\alpha$-detected SFACT objects.  This subsample includes a mixture of extended  galaxies where the fiber was placed on the center of the galaxy (magenta triangles), \ion{H}{2} regions located within these galaxies (typically the brightest emission knot; red circles), and the less extended, more compact ELGs (blue squares).  The solid curve represents a locus of high-excitation stellar photo-ionization models from \citet{DE86}, while the dashed line is the empirical demarkation line between AGN and star-forming galaxies proposed by \citet{kauff03}.  Only 56 of the 108 H$\alpha$-selected objects had all four of the lines necessary to form these two line ratios detected above the sensitivity threshold of the automated measurement software used by SFACT (see SFACT3).

The blue squares plotted in Figure~\ref{fig:diagplot1} include a fairly diverse set of objects.  They include many low-luminosity, low-metallicity systems plotted in the upper left portion of the diagram.   The objects in the lower right are mainly the more metal rich, luminous star-forming galaxies and their associated \ion{H}{2} regions (magenta triangles and red circles).  Two of the ELGs are located just above the \citet{kauff03} demarkation line, and may well be low-ionization nuclear emission region (LINER) AGNs.   Based on the combination of Figures~\ref{fig:absmaghist} and~\ref{fig:diagplot1}, we conclude that the ELGs represented by the H$\alpha$-selected portion of SFACT appear to span the full range of star-forming galaxies.

The spectral-line diagnostic diagram for the [\ion{O}{3}]-selected ELGs is shown in Figure~\ref{fig:diagplot2}, which plots the logarithm of [\ion{O}{3}]/H$\beta$ against the logarithm of [\ion{O}{2}]/[\ion{O}{3}].  As is the case with Figure~\ref{fig:diagplot1}, many of the SFACT galaxies detected via their [\ion{O}{3}] lines are not included in the figure because one or more of the necessary lines was not measured.  Only 70 of the 178 galaxies in this subsample have both line ratios available.   The dashed line represents the trend line shown in Figure 1 of \citet{bpt}, which is fit to emission-line ratios of approximately solar metallicity \ion{H}{2} regions and planetary nebulae.

The [\ion{O}{3}]-selected star-forming galaxies are located well above the BPT trend line.  This is to be expected, since their lower luminosities (Figure~\ref{fig:absmaghist}) coupled with the well-known trends from luminosity-metallicity relations \citep[LZRs, e.g.,][]{hirschauer2018} imply that these ELGs have, on average, substantially sub-solar abundances.  These lower metal abundances result in higher [\ion{O}{3}]/H$\beta$ ratios for a given value of the excitation ([\ion{O}{2}]/[\ion{O}{3}]).   If these galaxies could be plotted in Figure~\ref{fig:diagplot1} they would tend to be found in the upper left portion of the  star-forming galaxy sequence.

A single Seyfert 2 galaxy is also plotted in Figure~\ref{fig:diagplot2}.   It is located well above the BPT trend line and is clearly separated from the star-forming galaxies, consistent with expectations.  A total of three SFACT objects in the three pilot-study fields were provisionally classified at Seyfert 2s based on the appearance of their spectra (i.e., high [\ion{O}{3}]/H$\beta$ ratios).   However, the other two currently lack measurements of the [\ion{O}{2}] doublet, preventing us from including them in the figure.

\subsubsection{Star-Formation Rates}

\begin{figure}
\centering
\includegraphics[width=3.35in]{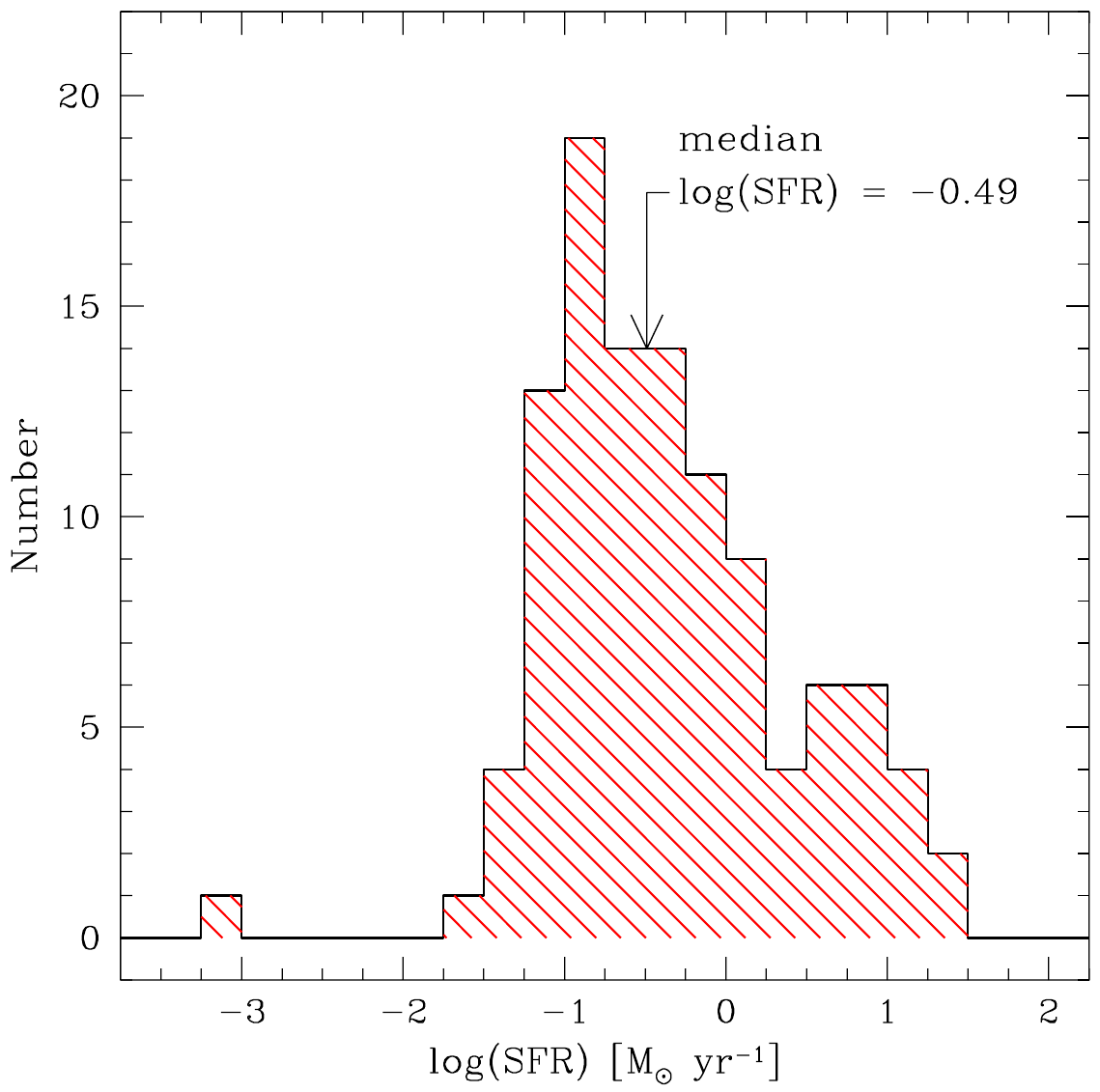}
\caption{Histogram showing the distribution of the star-formation rates for the H$\alpha$-detected SFACT ELGs in the three pilot-study fields (n = 108).  The object that stands out as having a low SFR is the sole H$\alpha$ detection in the NB2 filter.}
\label{fig:sfrhist}
\end{figure}

We computed H$\alpha$ star-formation rates (SFRs) for the galaxies in the lowest redshift windows that are detected by H$\alpha$.   We use the NB fluxes measured from our survey images for this purpose, since they are more accurately determined than the spectroscopic line fluxes and capture 100\% of the emission-line flux for our extended objects (which applies to essentially all of the H$\alpha$-detected sample).   For galaxies with multiple \ion{H}{2} regions (e.g., Figure~\ref{fig:images1}, top row), we compute the global SFR using the integrated line emission from the full galaxy.  The H$\alpha$ fluxes are corrected for the presence of [\ion{N}{2}] emission within the filter bandpass.   They are also corrected for absorption using the Balmer decrement (f(H$\alpha$)/f(H$\beta$)) when this quantity is measured in our spectra (as is the case for most H$\alpha$-detected sources).   We use the corrected H$\alpha$ flux and the redshift-determined distance to determine the H$\alpha$ luminosity, then utilize the standard \citet{kennicutt1998} relation (SFR [M$_\odot$ yr$^{-1}$] = 7.9 $\times$ 10$^{-42}$$\cdot$L(H$\alpha$)) to estimate the SFR.

The distribution of our derived SFRs is shown in Figure~\ref{fig:sfrhist}.   Despite the small size of the current sample and the limited survey volume of the three pilot-study fields, the SFACT sample is seen to be providing a robust measurement of the SFR distribution at the high end (log(SFR) $>$ 0.5).  The bulk of the sample is found to have log(SFR) values between $-$1.5 and 1.5 (0.03 $<$ SFR $<$ 30 M$_\odot$ yr$^{-1}$), with a median SFR value of 0.32 M$_\odot$ yr$^{-1}$.  The lack of detections at lower SFRs can be readily understood by referring to Figure~\ref{fig:zhist}.  Essentially all of the H$\alpha$ detections are made in the NB1 and NB3 filters.  Once again, the one exception is SFF01-NB2-B19198, the sole H$\alpha$ detection in NB2.  This galaxy stands out in Figure~\ref{fig:sfrhist} as having by far the lowest SFR of any object in the sample.

The current sample of SFRs is presented to illustrate the sensitivity of the SFACT survey and to establish the range over which the SFR measurements can be robustly determined.  Future papers in this series will explore the star-formation characteristics of much larger samples of SFACT galaxies, including the ELGs detected in the higher redshift windows.

\section{Comparison with Previous Narrowband Surveys} 

We compare the characteristics of the SFACT survey galaxies with a number of representative and recent NB surveys with the goal of helping to place the current survey into context.   We focus this comparison on three surveys that sample the ELG population in the same redshift range covered by SFACT: \citet{Stroe2015}, miniJPAS \citep{miniJPAS}, and LAGER \citep{LAGER20}.  We stress that each of the surveys being discussed was designed and carried out with a specific goal (or set of goals) in mind.  Each of the survey methodologies has obvious merit, and the resulting galaxy samples accomplished the goals set for the projects.  The aim of this comparison is not to rank the surveys in any way.   Rather, the purpose of this section is to allow the reader to better visualize the utility of the SFACT survey by comparing it directly to these other successful programs.

The \citet{Stroe2015} study presents a NB survey selecting galaxies via their H$\alpha$ emission.  The primary goals of this project were to securely measure the luminous end of the L$_{H\alpha}$ luminosity function at low redshift (z $\sim$ 0.20) and to quantify the level of cosmic variance present in small-area surveys.  In order to do this, the survey needed to cover a larger area on the sky, but it did not need to be particularly deep.  It utilized two NB filters that detected H$\alpha$ emitters in the redshift ranges 0.186$-$0.203 and 0.217$-$0.233.  The imaging was carried out on the 2.5 m Isaac Newton Telescope and used relatively short exposures (5 $\times$ 600 s).  Hence, the resulting galaxy sample is relatively shallow when compared to other surveys \citep[e.g.,][]{HIZELS12, HIZELS13}: the characteristic 50\% line flux completeness limit is $\sim$2 $\times$ 10$^{-15}$ erg s$^{-1}$ cm$^{-2}$.  The areal coverage of 12.8 deg$^2$ allowed the survey to detect sufficient numbers of lower redshift H$\alpha$-emitting galaxies to carry out the planned study.  The final sample of H$\alpha$-detected galaxies yielded 7.4 and 9.8 objects deg$^{-2}$ in the two NB filters.

The miniJPAS survey \citep{miniJPAS} is carried out in a manner which is quite different from most traditional NB surveys.  Rather than using one, or a few, NB filters to survey for emission lines at specific redshifts, miniJPAS utilizes a set of 54 filters with bandwidths of $\Delta\lambda$ $\sim$ 145 \AA.  These filters overlap each other so as to provide continuous wavelength coverage from 3800 $-$ 9100 \AA.  The miniJPAS survey, which covers 1 deg$^2$ overlapping the AEGIS \citep{aegis} field, is a fore-runner of the J-PAS survey, which will eventually cover $\sim$8000 deg$^2$.  While no emission-line flux completeness limits are provided, the sample appears to be 50\% complete for objects with a S/N of 5 at r $\sim$ 21.0.  The survey detected a total of 2154 potential ELGs, most of which are expected to be H$\alpha$-detections in the redshift range z = 0.00 $-$ 0.35.    Of these, 255 had sufficient S/N to allow for the determination of the emission line ratios [\ion{O}{3}]/H$\beta$ and [\ion{N}{2}]/H$\alpha$ with an uncertainty of 0.2 dex.   These detection rates are for the full redshift range specified above.  In order to directly compare these numbers with the NB surveys discussed here, we need to adjust for the limited bandpasses employed by the other surveys.  Adopting a characteristic filter width of $\Delta\lambda$ = 100 \AA\ results in a redshift coverage of $\Delta$z $\sim$ 0.016 near the middle of the miniJPAS redshift range.  This results in {\it approximate} detection rates of 98 ELGs deg$^{-2}$ per NB filter for all candidates, and 12 ELGs deg$^{-2}$ per NB filter for the high-quality subsample.

The Lyman Alpha Galaxies at Epoch of Reionization (LAGER) \citep{LAGER20} survey is a deep NB program being carried out with DECam on the CTIO Blanco 4.0 m telescope.  As the name implies, a key focus of the project is to detect Ly$\alpha$-emitting galaxies at high redshift (z $\sim$ 6.93).  In addition, the survey detected large numbers of ELGs via their H$\alpha$ (z = 0.47), [\ion{O}{3}] (z = 0.93), and  [\ion{O}{2}] (z = 1.59) emission lines.  The total integration time allotted to the single DECam field (FOV = 3.0 deg$^2$) with the LAGER NB filter ($\lambda_{cent}$ = 9640 \AA, $\Delta\lambda$ = 92 \AA) was 47.25 h, which resulted in extremely deep coverage.  Focusing on the H$\alpha$ detections to allow the best comparison with the other surveys, LAGER catalogued 1577 candidate ELGs with a 50\% line flux completeness limit of $\sim$2.5 $\times$ 10$^{-17}$ erg s$^{-1}$ cm$^{-2}$.  The detection rate of 526 H$\alpha$-detected ELGs deg$^{-2}$ in the single NB filter is extremely impressive.   The single field presented in \citet{LAGER20} overlaps the COSMOS field; the full LAGER survey proposes to cover a total of 8 fields and 24 deg$^2$.

As presented earlier in this paper, the SFACT survey has an approximate 50\% line flux completeness limit of $\sim$2 $\times$ 10$^{-16}$ erg s$^{-1}$ cm$^{-2}$, based on the three pilot-study fields.    This places SFACT midway between the \citet{Stroe2015} and \citet{LAGER20} surveys: the SFACT line flux completeness limit is approximately 10 times fainter than the \citet{Stroe2015} survey but is only 10\% as faint as the corresponding limit for the \citet{LAGER20} survey.   Based on the detection of 533 ELG candidates, the surface density of SFACT-detected galaxies is 355.3 ELGs deg$^{-2}$ across all filters, and 132.8 ELGs deg$^{-2}$ per NB filter.   This latter number accounts for the fact that there are essentially no H$\alpha$ detections for SFACT in the NB2 filter.  To better compare with the other surveys, we utilize our spectral data to derive detection rates of 50.5 deg$^{-2}$ for the H$\alpha$-detected ELGs in NB3 (z range of 0.129 $-$ 0.144) and an average of 42.0 ELGs deg$^{-2}$ for each of the three [\ion{O}{3}] redshift windows.  The SFACT detection rates are seen to be intermediate between those of the \citet{Stroe2015} and miniJPAS surveys on the one hand, and the LAGER survey on the other.

A key difference between SFACT and all of the surveys described above is the fact that SFACT has been designed from the start to include spectroscopic  follow-up of the sources detected.   The others rely on the use of photometric redshifts to indicate which emission line is detected in the NB filter.   While this approach works well on average, it is certainly not fool proof.   Furthermore, the lack of confirming spectra requires one to adopt the assumption that 100\% of the NB detections are in fact real ELGs.  Experience would suggest that no astronomical survey is perfect, and that some fraction of the objects selected as ELGs will be spurious.  Of course, having follow-up spectroscopy not only provides an absolute verification of the survey constituents, but it also allows for additional science applications.   While all of the comparison samples have been used successfully to estimate the star-formation rate density at the redshifts covered by their filter(s), the existence of spectroscopic follow-up will allow SFACT to also probe the evolution of the galaxy metal abundance with large, robust samples of ELGs and unambiguously identify rare objects like AGN and Green Pea galaxies (see \S 6).

\section{Applications of the SFACT Survey} 

In this section we describe a number of proposed applications for the SFACT survey.  This list is limited mainly by the interests of the current SFACT team members.  It is not meant to be exhaustive.   Rather, we hope to convey a glimpse of some of the exciting science outcomes expected from SFACT in the next several years.   Naturally, most of these projects will need to wait until we have obtained a more complete set of follow-up spectra.  As mentioned in the following section, however, this point is not necessarily that far off.  We are already working toward deriving preliminary results for a number of these applications based on the partial ELG samples currently in hand.

\subsection{Evolution of the Star-Formation Rate Density to z = 1 and Beyond}

The measurement of the star-formation rate density (SFRD) from the local universe to z =1 and beyond represents one of the major planned applications of SFACT, and the survey name was chosen as an epithetical reference to it.  The design of the SFACT survey should yield optimal results for the  measurement of the SFRD across this redshift range.  As specified in \S 3.3, we expect to have between 800 and 1200 emission-line-selected galaxies in each of our redshift windows, and these numbers are consistent with our results from the pilot-study fields presented in Figure~\ref{fig:zhist}, at least for the H$\alpha$ and [\ion{O}{3}]-detected objects.   These large samples will provide for robust estimates of the SFRD in each redshift window, and our field selection method will naturally account for cosmic variance.

 It is worth stressing that the sensitivity of the SFACT survey results in a comprehensive sample of star-forming galaxies within each redshift window.   This point is made evident by reference to Figure~\ref{fig:absmaghist}, which shows that all three primary lines detect galaxy samples that peak at or below the knee in the galaxy luminosity function (M$^*_g$ = $-$20.1; \citet{blanton2003}).   For example, the upper panel of Figure~\ref{fig:absmaghist} shows that the distribution of H$\alpha$-detected ELGs in the three pilot-study fields includes galaxies with g-band absolute magnitudes from M$_g$ = $-$21 (i.e., more luminous than M$^*_g$) to M$_g$ = $-$12.  The histogram suggests that the survey is fairly complete to M$_g$ = $-$16 for the H$\alpha$ detections.   Similarly, the middle histogram in Figure~\ref{fig:absmaghist} indicates the [\ion{O}{3}]-selected sample is fairly complete to M$_g$ = $-$17.  Even the galaxies detected via the [\ion{O}{2}] doublet probe galaxies with luminosities well below M$^*_g$.  This point is further solidified by examination of the SFR histogram shown in Figure~\ref{fig:sfrhist}.  The peak in the  distribution at SFR $\sim$ 0.1 M$_\odot$ yr$^{-1}$ corresponds to an H$\alpha$ luminosity of $\sim$1.3 $\times$ 10$^{40}$ erg s$^{-1}$, more than an order of magnitude below L$_{H\alpha}^{*}$ at these redshifts \citep{Stroe2015}.  Hence, the determination of the SFRD using SFACT galaxies will be based on large {\it and} comprehensive samples.  
 
Our current set of NB filters allows us to probe the star-forming galaxy population to z = 1.   In the near future we hope to add two additional filters (see Table~\ref{tab:filters}) that will detect galaxies via H$\alpha$ to z = 0.40, via [\ion{O}{3}] to z = 0.83, and via [\ion{O}{2}] to z = 1.46.  When the survey is complete, we expect SFACT to yield accurate SFRDs for each of redshift windows specified in Table~\ref{tab:filters}.  These measurements should provide ``hard points" in the distribution of SFRDs out to z $\sim$ 1.5. 

\subsection{Characterization of the Strong-Lined AGN population}

While the detection of star-forming galaxies is a particular strength of the SFACT survey method, our sample of ELGs will also include many AGNs.   These include Seyfert galaxies, LINERS, and QSOs.  While the number of AGN detected in our pilot-study fields is small relative to the star-forming galaxy population, when integrated over all survey fields and redshift windows we expect to detect them in substantial numbers, particularly in the H$\alpha$ and 
[\ion{O}{3}]-selected portions of the sample.  

Two specific areas of research that are of interest to the SFACT team members and that can be effectively explored with the survey data are the demographics of AGN at intermediate redshifts (z = 0.2 to 0.9) and the evolution of AGN metallicities with redshift.   Both topics are relatively under-explored but will be ripe for further study using SFACT.  For example, the survey will be able to be used to test whether the number density of AGNs increases in lock step with the density of star-formation activity with look-back time by measuring both using the same survey and within the same volumes of space.

There has been renewed interest in the determination of the metal abundances of AGN \citep[e.g.,][]{dors2020, flury2020, carvalho2020}.   Most of this work has focused on estimating the abundances of Seyfert 2 galaxies in the local universe.  However, a recent study that includes Seyfert 2 galaxies out to z = 0.4 suggests that higher redshift AGN possess lower metallicities than their low-z counterparts \citep{syabun}.   We expect to be able to probe the metal abundances of Seyfert 2 galaxies to redshifts approaching z = 0.9, to extend this result to higher redshifts and with larger samples of AGN.

\subsection{Demographics of Dwarf Star-forming Galaxies to z = 0.5}

It is clear from Figure~\ref{fig:absmaghist} that the SFACT selection method strongly favors the detection of intermediate- and low-luminosity galaxies, particularly within the H$\alpha$ and [\ion{O}{3}]-detected portions of the survey.  This result comes about for a number of reasons, including the metallicity-related effect on the strength of the [\ion{O}{3}] doublet mentioned in \S 4.3.1.  Other factors include the sensitivity of the survey to faint sources, the increased contrast between strong star-forming knots and the underlying continuum in low-luminosity systems, and the simple fact that dwarf galaxies -- star-forming or otherwise -- are more common in the universe than more luminous galaxies.   SFACT can readily detect galaxies with absolute magnitudes of M$_g$ = $-$15 to z $\sim$ 0.15 via the H$\alpha$ line and M$_g$ = $-$16 out to z $\sim$ 0.50 via the [\ion{O}{3}] line.

These characteristics of the survey will allow us to probe the properties and demographics of dwarf star-forming systems to substantial redshifts.  In particular, SFACT will detect large samples of BCDs in all of the redshift windows below z = 0.50.  This will allow for the unprecedented opportunity to study this important class of star-forming galaxy to cosmologically significant distances.  With sufficient follow-up spectral data, it will be possible to probe for  redshift dependences in the metal abundances of the BCDs.  We also expect to be able to constrain the evolution of the dwarf star-forming galaxy population from z = 0.5 to today.

\subsection{Environments of Star-Forming Galaxies and AGNs}

As described in \S 3.3, many of the SFACT survey fields coincide with fields that contain a previously known Green Pea galaxy.  \citet{brunker2022} has carried out a redshift survey in these fields in order to study the galactic environments that Green Peas are located in.  We plan to build upon this previous work by continuing to obtain redshifts of galaxies in these fields as part of the SFACT follow-up spectroscopy campaign.  Every multi-fiber configuration observed with Hydra includes non-SFACT galaxies selected from SDSS in any extra fibers.   Since most of the SFACT fields require 3 to 5 Hydra configurations to observe all of the ELG candidates, these extra fibers will yield 50 to 100 additional field-galaxy redshifts in each field.   In addition, we have plans to carry out a more focused redshift survey of field galaxies in several of the SFACT fields, with the goal of acquiring a fairly complete redshift sample for galaxies to g $\sim$ 21.  These extra redshifts for faint SDSS galaxies located within our survey fields will be in addition to the existing SDSS redshift survey data \citep{strauss2002}.   This will allow us to probe in detail the distribution of galaxies out to z $\sim$ 0.5.

The science driver for obtaining these redshifts is to allow us to study the environments of the SFACT star-forming galaxies and AGN.   With a typical yield of $\sim$150 ELG candidates per field, the SFACT survey provides an excellent opportunity for studying the effect that environment plays on driving activity in galaxies.  The planned redshift survey of field galaxies located in and around the SFACT fields will be deep enough to probe environmental impact using all of the H$\alpha$ and [\ion{O}{3}]-detected objects.  The rich sample of ELGs in each SFACT field will provide a statistically meaningful ensemble of galaxies with which to probe the impact of local environment of  both star-formation and AGN activity.

\subsection{Evolution of Galaxy Abundances to z = 0.9}

The measurement of the metal abundances in galaxies is a key tool for understanding how they evolve with time.   As our picture of galaxy evolution has taken shape, astronomers have come to realize that many physical processes -- beyond basic star evolution -- can affect the chemical enrichment of a given galaxy.  For example, galaxy mergers with metal-poor but gas-rich dwarfs or the infall of pristine gas can lower the overall metallicity of a galaxy.  These effects are likely to be dependent on the local environment of the system.   The outflow of metal-rich ejecta from supernovae can likewise lower the measured abundance of a galaxy below the level expected based on its time-integrated star formation history.  This process will be dependent on the overall mass of the galaxy.   Disentangling all of the relevant processes to arrive at a more complete picture of galaxy chemical evolution from the observational side requires large, comprehensive samples of galaxies with metal abundances.

The SFACT survey holds much promise for providing the type of galaxy sample necessary to make substantial progress in probing the metal abundances of large samples of star-forming galaxies to cosmologically relevant distances.   With many hundreds of galaxies with measured abundances within each redshift window, SFACT will be able to deliver very focused views of the metallicities of galaxies.   In particular, we envision constructing luminosity-metallicity and mass-metallicity relations (LZRs and MZRs, respectively) for each redshift window.  We expect to be able to robustly map out the redshift evolution of the LZR and MZR to the redshift limit of our data. 

Our current round of follow-up spectroscopy should be adequate for providing abundance estimates for the majority of star-forming galaxies in the H$\alpha$ and [\ion{O}{3}]-selected subsamples, where the necessary nebular diagnostic lines are present in our data.   However, spectra that reach to redder wavelengths will be necessary to yield metallicity estimates for any of the [\ion{O}{2}]-selected ELGs (see \S 7).  With these new red-spectral-range data we expect to be able to derive metallicities for galaxies detected via the [\ion{O}{2}] doublet discovered in NB1 and NB2.  This will push our analysis of the LZRs and MZRs out to z $\sim$ 0.88.  Complete metallicity studies involving the very faintest of the SFACT objects will likely require additional spectroscopy using larger telescopes.

\subsection{Detection of Rare Objects}

\begin{figure}
\centering
\includegraphics[width=3.35in]{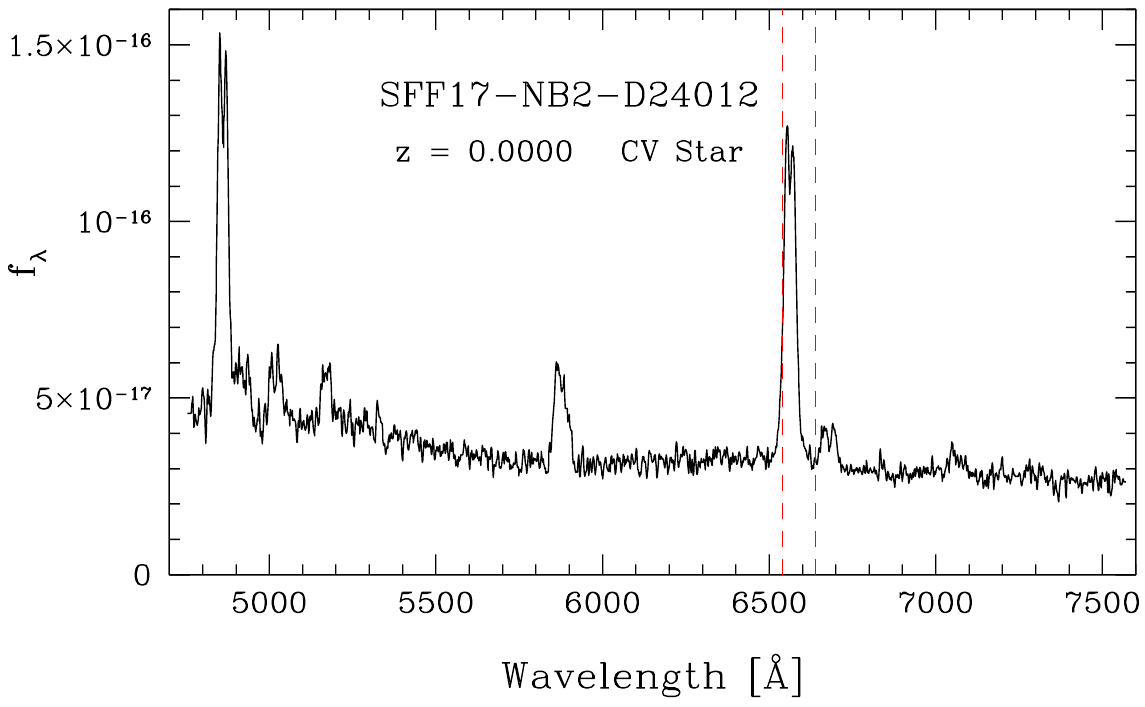}
\caption{Spectrum of the cataclysmic variable (CV) star SFF17-NB2-D24012.  This object was previously identified in a catalog of suspected CVs by \citet{drake2014}.   It was detected in our survey due to the strong H$\alpha$ emission located in NB2.  Other lines visible in the spectrum include H$\beta$ as well as five Helium lines: \ion{He}{1} 4921, 5015, 5876, 6678, 7065.  The lines are double peaked and broad, indicative of emission from a rapidly rotating disk.}
\label{fig:spec_cv}
\end{figure}

One of the more enjoyable and unpredictable aspects of carrying out astronomical surveys of this type is the fact that one commonly encounters unusual objects.  In extreme cases, one might even discover an entirely new class of objects.  While the SFACT survey is still in its early days, we have already come across a number of interesting objects.  For example, we have ``discovered" a cataclysmic variable (CV) star in our early survey data that had previously been identified as a suspected CV candidate by  \citet{drake2014}.  Its spectrum is shown in Figure~\ref{fig:spec_cv}.  More to the point of our ELG survey, we expect to detect a number of rare and potentially very interesting objects such as Green Pea galaxies and extremely metal-poor (XMP) dwarf galaxies.  

On face value, our survey method may not appear to be the best approach for discovering rare objects.   As already discussed, a primary drawback of the NB survey technique is the relatively small volumes covered with each pointing.  The SFACT survey methodology mitigates this problem to some extent by using somewhat broader filters than have traditionally been used in the past ($\sim$90 \AA\ rather than $\sim$50-60 \AA).  These broader filters, coupled with the large FOV of the ODI camera, result in reasonably large volumes for each redshift window.  For example, the effective survey volume for a single field with the NB1 filter detecting ELGs via the [\ion{O}{3}]$\lambda$5007 line (redshift range of 0.379 to 0.397) is of order 80,000 Mpc$^3$.   If one then multiplies this by the three principle emission lines detected in each pointing, by the three NB filters currently used by the survey, and by the 50-60 planned survey fields, the total effective volume of the survey rises to the level of tens of millions of cubic megaparsecs.   Hence, our expectation is that many dozens of interesting objects will be found during the course of the survey.

We highlight a few categories of rare objects below.

\subsubsection{Green Peas and Blueberries}

\begin{figure}
\centering
\includegraphics[width=3.35in]{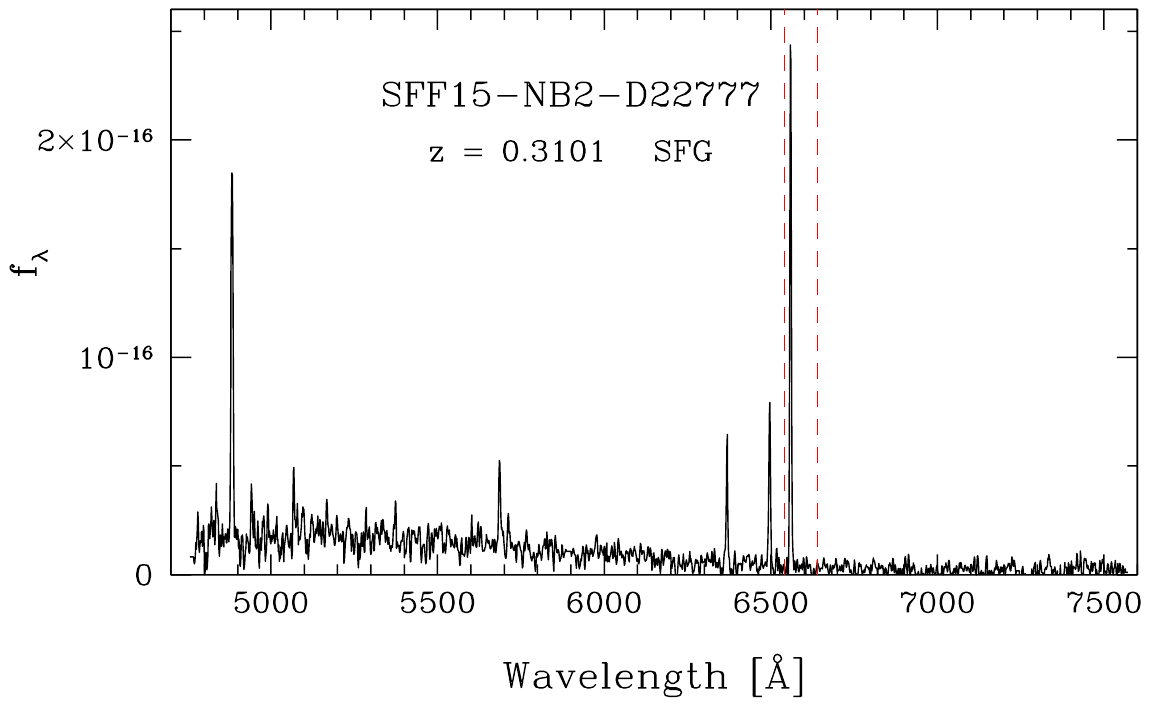}
\caption{Spectrum of a candidate Green Pea galaxy SFF15-NB2-D22777.  The combination of high equivalent width [\ion{O}{3}] emission lines and high luminonsity (M$_g$ = $-$19.3) are typical of the Green Peas.}
\label{fig:spec_gp}
\end{figure}

The Green Pea \citep{gp} and Blueberry \citep{yang2017} galaxies are among the most extreme star-forming galaxies known.  Their common feature is the presence of very strong [\ion{O}{3}]$\lambda\lambda$5007,4959 emission.   The original samples of both types of galaxies were created using BB colors that were sensitive to high-equivalent-width [\ion{O}{3}] emission.   Due to this selection method, the two sets of galaxies are only detected in limited redshift ranges that are not overlapping.  Currently very little is known about the redshift evolution of either the Green Pea or Blueberry populations, or how their properties compare with less extreme star-forming galaxies.   Do the Green Peas and Blueberries form a continuum of extreme objects, or are they distinctly different types of systems?  

Emission-line-selected samples \citep[e.g.,][]{brunker2020} have shown that these types of systems can be identified over a much broader range of redshifts.   We expect that the [\ion{O}{3}]-selected subsample of the SFACT survey should be particularly effective at detecting both types of galaxies over an extended redshift range.  SFACT should also be sensitive to lower-redshift versions in the H$\alpha$-detected portion of the survey.  The spectrum of an example SFACT Green Pea candidate is shown in Figure~\ref{fig:spec_gp}.   This galaxy, SFF15-NB2-D22777, has a redshift of z = 0.3101 and a g-band absolute magnitude of M$_g$ = $-$19.3.   The high luminosity coupled with the large equivalent widths of the [\ion{O}{3}] line (EW$_{5007}$ $\sim$ 600 \AA) are what identify this galaxy as a Green Pea candidate.   As the survey progresses, we expect to build up a large population of both Green Peas and Blueberries.   This will allow us to carry out a complete study of their demographics and better place them into context with the broader population of star-forming galaxies.

\subsubsection{XMP galaxies}

\begin{figure}
\centering
\includegraphics[width=3.35in]{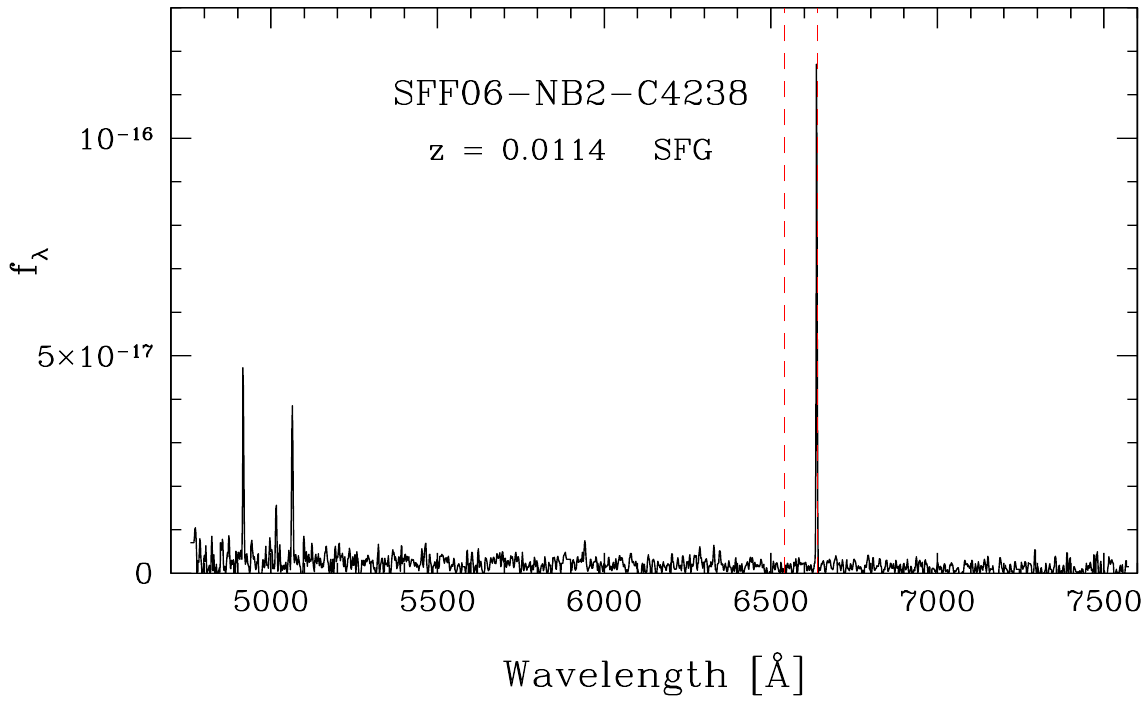}
\caption{Spectrum of the metal-poor dwarf galaxy SFF06-NB2-C4238.  It has an estimated oxygen abundance log(O/H)+12 of 7.15 $\pm$ 0.10.}
\label{fig:spec_xmp}
\end{figure}

The discovery and subsequent study of XMP galaxies has a long and interesting history \citep[e.g., see][for a recent review]{mcquinn2020}.   Due to the nature of the LZR, the most extremely-low-abundance systems should also be extremely-low-luminosity, low-mass systems.  This has made their detection a challenge.   Despite a focused effort to discover more of these galaxies (e.g., see the list of previous surveys mentioned in \S 1), the number of truly XMP galaxies (e.g., log(O/H)+12 $<$ 7.3) remains modest.

The discovery of XMP galaxies was {\it not} among the list of expected SFACT survey results when the survey began.   The primary reason for this has already been mentioned in \S 3.2 and verified in \S 4.2.2: the effective survey volume for the H$\alpha$-detected NB2 redshift window is quite small.  We  expect that this is the only one of the current survey redshift windows where we would be sensitive to the very low-luminosity XMP objects.   For example, the next-lowest redshift window is the H$\alpha$-selected NB1 region, which covers the redshifts z = 0.052 - 0.066.   The XMP galaxies located at these redshifts are likely to be below our detection threshold.

Despite this gloomy outlook, we can report the discovery of at least a few XMP candidates in the early survey data.  We illustrate one example in Figure~\ref{fig:spec_xmp}.  The object is SFF06-NB2-C4338 (which is {\it not} from one of the three pilot-study fields).  It has a g-band magnitude of 23.05 and a redshift of z = 0.01144 (distance = 53.2 Mpc).   The inferred absolute magnitude is M$_g$ = $-$10.6.   The spectrum exhibits an [\ion{O}{3}]$\lambda$5007 line that is weaker than H$\beta$, plus strong H$\alpha$ but no detection of [\ion{N}{2}]$\lambda$6583, both characteristics of XMP systems.  A spectrum covering bluer wavelengths (not shown) reveals a weak detection of the [\ion{O}{2}] doublet.  This has allowed us to generate a preliminary estimate of the metal abundance using the R23-O23 method \citep[e.g.,][]{mcgaugh} which yields log(O/H)+12 = 7.15 $\pm$ 0.10.  More details, as well as results from other low metallicity objects, will be presented in forthcoming papers.

\section{Status and Future Plans} 

The SFACT survey has been acquiring imaging data with the current set of three NB filters since the Fall 2018 semester.  Despite a significant loss of observing time due to technical problems, weather, and health safety issues (i.e., WIYN was closed for most of 2020 due to the Covid-19 pandemic), our team has been making steady progress toward meeting the overall survey goals outlined in \S 3.3.  In the current section we summarize the status of our survey observations.   We also highlight future plans and possible new directions for SFACT.

As of the end of the Spring 2022 semester, we have reached the $\sim$75\% level toward our goal of observing 50-60 fields.  The SFACT team has acquired complete NB and BB imaging for 43 fields.  Of these, 24 are Fall fields and 19 are located in the Spring sky.   Processing has been completed on 41 of these fields, resulting in the detection of $\sim$5500 ELG candidates.  Our current analysis efforts are focusing on completing the processing for the remaining fields for which data exist, in order to initiate spectroscopic follow-up observations.   Our current plans are for the acquisition of imaging survey data with the three existing NB filters to continue through the 2023 observing seasons.  We expect to reach our target goal for the number of fields observed at that point.

Naturally, the acquisition of follow-up spectroscopy lags the imaging portion of the survey.  However, for the past several semesters we have utilized roughly 50\% of our observing time to obtain these extremely important data.  Most of our fully-processed fields have at least some follow-up spectroscopy, ranging between $\sim$25\% and $\sim$90\% coverage.  As time goes on, an increasing fraction of our observing efforts will shift to  spectroscopic confirmation.  However, we already have a solid start on this phase of the project.  For example, there are 17 Fall-semester fields that have been processed through the stage of ELG candidate selection that have significant amounts of follow-up spectra.  These fields include a total of $\sim$2300 ELG candidates, of which 55\% already possess follow-up spectra.

Two major modifications to our observing procedures are planned for the future.   First, as outlined in \S 3.2, we plan to add additional NB filters to the survey.  Both of these new filters are located in gaps in the telluric OH spectrum, providing relatively dark spectral windows suitable for our survey work.   These filters serve three important roles for the survey.  First, they will create additional redshift windows, filling in gaps in our current redshift coverage.  For example, NB812 provides key redshift windows at z $\sim$ 0.24 (H$\alpha$ detections) and z $\sim$ 0.62 ([\ion{O}{3}] detections).  See Figures~\ref{fig:filters} and~\ref{fig:zhist} for more details.  Second, they significantly extend the redshift coverage over which we will be able to detect star-forming ELGs (up to as high as z $\sim$ 1.46).  Finally, the addition of the NB912 filter allows for the detection of objects within the {\it same redshift window} via both the H$\alpha$ and [\ion{O}{3}]$\lambda$5007 lines.  The latter are detected using the current NB1 filter.  This overlapping selection capability will provide the opportunity to directly compare the ELG populations found in the same volumes of space via these two distinctly different lines, and to robustly calibrate the SFRs found for the [\ion{O}{3}]-selected galaxies, firmly placing them on the same scale as the H$\alpha$ SFR measurements.

A second change to our survey methodology that will be implemented in the coming semesters will be the acquisition of spectra at longer wavelengths.  As seen in Figure~\ref{fig:zhist}, the majority of the SFACT ELGs are detected via their [\ion{O}{3}]  and [\ion{O}{2}] lines.  However, our current set of follow-up spectra only cover the wavelength range of 4700 -- 7600 \AA.   With this spectral coverage we are not able to observe the important spectral region near H$\alpha$ for the [\ion{O}{3}]-detected SFACT galaxies, missing the important [\ion{N}{2}]$\lambda\lambda$6583,6548 and [\ion{S}{2}]$\lambda\lambda$6731,6716 diagnostic emission lines.  The situation is even worse with the [\ion{O}{2}]-detected galaxies, where our current spectral coverage tends to cover only the [\ion{O}{2}] doublet itself plus the [\ion{Ne}{3}]$\lambda$3869 line.  Once our first round of follow-up spectroscopy is complete, we plan to pursue a second round for our [\ion{O}{3}]-  and [\ion{O}{2}]-detected ELGs.  Naturally, once we start observing SFACT fields using the proposed new NB812 and NB912 filters we will need to observe further into the red simply to obtain verification spectra.   Hence we expect to carry out the second-pass red spectroscopic campaign simultaneously with the first-pass follow-up spectra for objects detected in the new longer-wavelength filters.

\section{Summary \& Conclusions}  

We present a description of the new NB imaging  survey SFACT: Star Formation Across Cosmic Time.  SFACT is a long-term program being carried out on the WIYN 3.5 m telescope using both the ODI wide-field imaging camera and the Hydra multi-fiber positioner for spectroscopic follow-up.  In addition, we present preliminary results from newly discovered SFACT objects detected in three pilot-study fields.

The imaging portion of the survey utilizes the ODI camera, which has a field-of-view of 48 $\times$ 40 arcmin ($\sim$0.53 deg$^2$).   Currently, three specially designed NB filters are used to detect ELGs at a range of redshifts up to z $\sim$ 1.0 and QSOs to z $\sim$ 5.2.  Future expansion of the survey by adding additional NB filters is planned.   In addition to the NB images, survey fields are imaged through {\it gri} BB filters.  The latter images provide deep calibrated photometry as well as providing the necessary continuum subtraction for our NB data.  Our overall survey plan calls for observing between 50 and 60 survey fields, which will result in the detection of on the order of 1000 ELGs in each redshift window covered by the survey.   

The three pilot-study fields yielded a total of 533 emission-line sources from the imaging survey.  This represents a surface density of 355 objects deg$^{-2}$, which is entirely consistent with the expectations based on our projections for the survey depth.  The median r-band magnitude of the SFACT sources is 22.51, and the faintest objects have r $\sim$ 25.   The median NB line flux measured for our ELG candidates is 2.97 $\times$ 10$^{-16}$ erg s$^{-1}$ cm$^{-2}$, and the limiting flux is 1.01 $\times$ 10$^{-16}$ erg s$^{-1}$ cm$^{-2}$.   We find a good distribution of detections between the three NB survey filters.  Overall we interpret the results from the pilot study as a strong verification of the validity of our survey method.

The second component of the SFACT survey entails the acquisition of spectra of the ELG candidates.  These spectra are necessary for confirming the candidate ELGs detected in the imaging data, for determining which line was present in the NB filter, and for determining redshifts which in turn allows us to derive important physical properties of our sample galaxies.  While the spectroscopic follow-up naturally must lag the imaging portion of the survey, we have nonetheless already acquired substantial amounts of spectra in the early days of the survey.   For the pilot-study fields, 453 out of the 533 ELG candidates (85.0\%) have follow-up spectra.  Of these, 415 are confirmed to be {\it bona fide} emission-line objects (91.6\%).  We use the results of the spectroscopy to determine the redshift distribution of the SFACT galaxies, and present preliminary looks at the luminosity and SFR distributions of our sample.

Two companion papers present our initial survey catalogs of extragalactic emission-line objects as well as the available follow-up spectroscopy for these sources.  \citet[][SFACT2]{sfact2} describes the ODI imaging observations of our three pilot-study fields, details the data processing and analysis steps applied to the data to yield our survey lists, and presents the first three lists of SFACT ELGs.  Numerous examples of newly discovered ELG candidates are shown to illustrate the range of objects discovered in SFACT.    \citet[][SFACT3]{sfact3} presents the results of our spectroscopic follow-up of the objects from the pilot-study fields.  After detailing our observing methods, data-processing steps, and emission-line-measurement procedures,  the paper tabulates all of the relevant spectral information.   Many example spectral plots are used to illustrate the types of ELGs detected in SFACT.

As described above, we have already acquired imaging and spectra data for a substantial number of additional survey fields.  We expect that we will be publishing survey lists and follow-up spectroscopy at regular intervals.   We are also open to sharing our survey lists prior to publication.  Colleagues who would like to work with our deep survey data are encouraged to contact us.

\acknowledgements

We gratefully acknowledge the longterm financial support provided by the College of Arts and Sciences at Indiana University for the operation of the WIYN Observatory.  Additional funds have been provided by the Department of Astronomy and the Office of the Vice Provost for Research at Indiana University to help support this project.   The authors express their appreciation to the anonymous referee who made a number of insightful suggestions that improved the quality of this paper.  The entire SFACT team wishes to thank the entire staff of the WIYN Observatory, whose dedication and hard work have made this survey possible.   In particular, we acknowledge the contributions of Daniel Harbeck, Wilson Liu, Susan Ridgeway, and Jayadev Rajagopal.  We also thank Ralf Kotulla (U. Wisconsin) for his development and continued support of the ODI image processing software (QuickReduce), and Arvid Gopu and Michael Young (Indiana U) for their support of the ODI Pipeline, Portal \& Archive.  Finally, we acknowledge the contributions made at various stages of this project by students in the Department of Astronomy at Indiana University who assisted with the data processing: Bryce Cousins, Anjali Dziarski, Sean Strunk, and John Theising.

\facilities{WIYN:3.5m}

\end{document}